\documentclass{aa}  
\usepackage{graphicx}
\usepackage{txfonts}
\usepackage{amsmath}	
\usepackage{amssymb}	
\usepackage{subfig}

\begin{document} 

\title{Accretion and jets in a low luminosity AGN: the nucleus of NGC~1052}


\author{S. Falocco
          \inst{1}\fnmsep\thanks{E-mail: falocco@kth.se}
          \and
          J. Larsson\inst{1}
          \and
          S. Nandi\inst{2}
}


\institute{
$^{1}$ KTH, Department of Physics, and the Oskar Klein Centre, AlbaNova, SE-106 91 Stockholm, Sweden \\
  $^{2}$ National Centre for Radio Astrophysics, TIFR, Pune University Campus, Post Bag 3, Pune 411 007, India
}

\date{Accepted XXX. Received YYY}

\abstract{}{We aim to determine the properties of the central region of NGC~1052 using  X-ray and radio data. NGC~1052 (z=0.005) has been investigated for decades in different energy bands and shows radio lobes and a low luminosity active galactic nucleus (LLAGN).
}{We use X-ray images from {\it Chandra} and radio images from Very Large Array (VLA) to explore the morphology of the central area. We also study the spectra of the nucleus and the surrounding region
  using observations from {\it Chandra} and {\it XMM-Newton}.}
{We find diffuse soft X-ray radiation and hotspots  along the radio lobes.
  The spectrum of the circum-nuclear region is well described by a thermal plasma (T$\sim 0.6$ keV) and a power law with  photon index $\Gamma\sim2.3$.
  The nucleus shows a hard power law ($\Gamma\sim1.4$) modified by complex absorption. A narrow iron K$\alpha$ line is also clearly detected in all observations, but there is no evidence for relativistic reflection.  }{
  
The extended emission is consistent with originating from extended jets and from jet-triggered shocks in the surrounding medium. The hard power-law emission from the nucleus and the lack of relativistic reflection supports the scenario of inefficient accretion in an Advection Dominated Accretion Flow (ADAF).
} 

\keywords{
X-rays: galaxies --  Galaxies: nuclei  --  Galaxies: jets  }

  \maketitle

\section{Introduction}

Low Ionisation Nuclear Emission Line Regions (LINER) are characterised by intense optical lines from low ionisation species \citep{heckman1980}. Although their optical properties define them as a class, there is not yet a consensus on their ionising mechanism. Indeed, the LINER class is highly heterogeneous, including sources with and without clear evidence of an Active Galactic Nucleus (AGN), as well as both actively star-forming and passive galaxies. Alternatives to photoionisation explain the line excitation as shock-heated regions of the Inter Stellar Medium (ISM) \citep{dopita1997,sugai2000}, strongly star forming regions \citep{armus1990,balmaverde2015} or diffuse ionised plasmas \citep{collins2001}. However, several LINER surveys have shown that the majority of sources are AGN powered by low-luminosity accretion onto SMBHs \citep{ho2008,terashima2002,gonzalez-martin2009b,hernandez-garcia2013,hernandez-garcia2014}. The evidence collected so far provides support for the scenario that these sources have ADAFs embedded in standard accretion discs, as described by \cite{narayan1998}. This picture has been adopted to explain most LINER as LLAGN \citep{ho2008}.
Since low and high ionisation states correspond to low and high accretion rates, LINER can be seen as the low accretion rate extension of Seyfert galaxies \citep{alexander2012}. LINER thus offer an opportunity to study SMBHs in the low accretion rate regime, which is thought to dominate in the local universe (e.g. \citealt{hickox2009, alonso-herrero2008}). 

The X-ray emission from LINER offers a powerful way of unveiling the presence of an AGN. Indeed, both spectral features and variability  support the scenario of SMBH accretion in LINER  \citep{gonzalez2009,hernandez-garcia2014,hernandez-garcia2016}.  One unambiguous evidence of AGN-like accretion is the narrow iron K$\alpha$ line at 6.4~keV. This line  is ubiquitous in AGN and originates from the outer parts of the accretion disk or regions further away, such as the torus or  Broad Line Region.  AGN commonly also exhibit a relativistically broadened iron line emitted from the innermost accretion disc, along with other reflection features such as the Compton hump above 10~keV (e.g., \citealt{garcia2014}). Such relativistic reflection is not expected to be observed in the ADAF scenario for LINER, as the inner accretion flow is then optically thin and the iron is fully ionised. A relativistic line has been reported in only one source (NGC~1052, \citealt{brenneman2009}), discussed further below.
The narrow iron line is instead rather common (e.g. 42 \% in the sample of \citealt{terashima2002}), offering evidence for the presence of AGN. 

The X-ray spectra of LINER often have significant contributions from a variety of components  in addition to the emission associated with accretion. Such components can be detected due to the typical low redshifts and intrinsic weakness of the nuclei. In soft X-rays, a mixture of radiation from supernova remnants and starburst regions are sometimes used to model the thermal X-ray continuum \citep{gonzalez2009}. Emission from supernovae and hot ($\sim10$~keV) gas from starburst activity can also contribute at hard X-rays \citep{terashima2002}. Some LINER also exhibit X-ray emission from jets \citep{hardcastle1999}, which can be be due to synchrotron emission or inverse Compton scattering (e.g. \citealt{worrall2009,hardcastle1999}, \citealt{tavecchio2000,celotti2001}). 

NGC~1052 is considered a LINER prototype. 
 It has been studied extensively in the literature (e.g.  \citealt{heckman1980} and \citealt{kadler2004})  and been included in numerous  X-ray surveys of LINER \citep{gonzalez-martin2009b,hernandez-garcia2013,hernandez-garcia2014,gonzalez-martin2014, gonzalez-martin2015, hernandez-garcia2016}. It is hosted by an elliptical galaxy located at a distance of 19.2~Mpc \citep{tully2013} with redshift  $z=0.005$ \citep{Jensen2003}, and its black hole mass has been estimated to  $1.5\times10^8$~M\textsubscript{\(\odot\)} \citep{woo2002}.  It  exhibits kpc-scale twin radio jets and associated lobes (e.g., \citealt{kadler2004}). On pc-scales the two jets are separated by a central gap that becomes smaller with increasing frequencies, which has been interpreted as absorption from a torus surrounding the nucleus \citep{kadler2004h, baczko2019}. There is also optical line emission aligned with the inner radio jet, presumably originating from shock-excited regions \citep{kadler2004,dopita2015}.

NGC~1052 was first resolved in X-rays in a short {\it Chandra} observation from 2001, which revealed soft X-ray emission associated with the kpc-scale radio jets \citep{kadler2004}. This extended emission was well described by a  $0.4-0.5$~keV thermal plasma and interpreted as originating from jet-triggered shocks. 
The broad-band X-ray emission from NGC~1052 is, however, dominated by the nucleus. Early studies of the X-ray spectrum revealed a flat power-law continuum affected by significant intrinsic absorption, as well as a narrow Fe line at 6.4~keV \citep{guainazzi1999,guainazzi2000}. The flat spectrum was interpreted as Bremsstrahlung from an ADAF.  A more detailed study  based on a {\it Suzaku} observation from 2007  confirmed the basic picture of an absorbed hard power-law, but also revealed the presence of a relativistically broadened Fe line \citep{brenneman2009}. However, no associated Compton reflection was detected, making an origin in the inner region of a standard accretion disk unlikely. \cite{brenneman2009} instead suggested that the broad line may have been produced near the base of the jet.

More recently, 
\cite{osorio2019} presented an extensive study of NGC~1052 using observations from {\it Chandra}, {\it XMM-Newton}, {\it Nustar}, and {\it Suzaku}. Their preferred model for the nucleus comprises an absorbed power law and reflection from distant material, while the spectrum of the circum-nuclear region was fitted with a very hard power law together with a thermal component, consistent with \cite{kadler2004}. \cite{osorio2019} also reported variability in the nucleus. 

From the current knowledge and previous work on NGC~1052, it is clear that open questions still remain. One of these concerns the nature and origin of the complex soft X-ray emission from the circum-nuclear region, including the possible role of star formation and the connection with the radio jets. The details of the circum-nuclear spectrum also affects the modelling of the nucleus, including the properties of the complex absorption,  spectral variability, and the possible presence of relativistic reflection. The latter point is especially interesting because it is not expected in an ADAF and because there have been discordant results regarding the presence of a broad iron line in the literature. 

This paper presents an analysis of one {\it Chandra} and four {\it XMM-Newton} observations of NGC~1052 with the purpose to explore these aspects in more detail.
The excellent spatial resolution of {\it Chandra} allows us to study the morphology of the central region, as well as the X-ray spectra
associated with the jets, the galaxy, and the nucleus. We also present the first comparison between the long {\it Chandra} observation of 2005 and the VLA observations.
The information obtained from {\it Chandra} is also used when analysing the {\it XMM-Newton} observations, which offer good-quality spectra but do not resolve the galaxy. This approach allows us to characterise the spectra of the nucleus in detail and to determine whether the variability is significant on time scales of years. We describe the observations in Section 2, present the analysis of images and spectra in Section 3, and discuss the results in Section 4. The conclusions are summarised in Section 5.

\section{Observations and data reduction}

The  {\it Chandra} and {\it XMM-Newton} observations analysed in this work are  listed in Table~\ref{tab:obs}.
 For comparison with the {\it Chandra} images, we also retrieved a radio image at 1.4 GHz from the VLA archive.

\begin{table*}
\centering
\caption{Details of the X-ray observations. Exp is the exposure time while exp$_{\rm clean}$ is the net exposure remaining after removing times affected by background flares from the instruments specified (MOS1+MOS2 or pn). Count rates are given for the $0.5-10$~keV energy range for the different instruments (MOS1+MOS2, pn and ACIS). Superscript 1  means that only MOS1 was used. In the case of {\it Chandra} ACIS, the count rate is for the annular extraction region at the nucleus.}
\label{tab:obs}
\begin{tabular}{lccrcccccc} 
 \hline
 &  &  &  & MOS &  pn    &   MOS   & pn & ACIS\\
Date	 & Telescope & OBSID & exp &   exp$_{\rm clean}$ &   exp$_{\rm clean}$  &   rate   & rate & rate\\
 &  &  & [ks]   & [ks]  & [ks]   &   [$\rm{cts~s^{-1}}$]   & [$\rm{cts~s^{-1}}$]   & [$\rm{cts~s^{-1}}$]\\
\hline
2001 Aug 15	 &   {\it XMM-Newton}       & 0093630101   & 16.3  & -     &  10.4   &  -           &     0.4  & - \\
2005 Sept 18        &   {\it Chandra}    & 5910        & 59.2  & -     & -       &    -   & -  &   0.037 \\
2006 Jan 12	 &   {\it XMM-Newton}       & 0306230101   & 54.9  & 100.0   & 39.8    & 0.16         &  0.5  & - \\
2009 Jan 14        &   {\it XMM-Newton}       & 0553300301  & 52.3  & 47.0$^{1}$    & 39.9    &  $0.2^{1}$     &  0.6  & -\\
2009 Aug 12        &   {\it XMM-Newton}       & 0553300401  & 59.0  & 110.0   & 41.8    &   0.18       & 0.6  & - \\
\hline
\end{tabular}
\end{table*}

\subsection{Chandra observation}
\label{sec:chandradata} 

There are two {\it Chandra} observations of NGC~1052, one from 2001 (observation Identification, OBSID 385) and one from 2005 (OBSID 5910). Both observations were performed with the ACIS-S instrument. We analysed only the 2005 observation because it has a significantly longer exposure time and because the first observation has already been presented in \cite{kadler2004}. The details of the 2005 observation are reported in Table \ref{tab:obs}. We note that contamination on the optical blocking filter (which affects {\it Chandra} data below $\sim 1$~keV) was small at the time of the observation and hence does not significantly affect our analysis \citep{contamination2018}.

The data were reduced using {\texttt CIAO} v4.9 \citep{fruscione2006}, with calibration files version 4.7.4.  We checked that there was no background flaring present in the observation. However, the nucleus is affected by pileup at a level of $\sim 10$~per cent (estimated using {\texttt PIMMS}\footnote{\url{http://heasarc.gsfc.nasa.gov/docs/software/tools/pimms.html}}). To account for this potential problem we extracted the spectrum of the nucleus from an annular region with inner and outer radii of 1$\arcsec$ and 6$\arcsec$, respectively.  The background was extracted in a source-free nearby circular region with a radius of 10$\arcsec$. The spectrum was binned to 25 counts per bin in order to allow for fitting using $\chi^2$ statistics. 

We also extracted spectra in two rectangular regions located to the east and west of the nucleus in order to analyse the emission from the jet and galaxy (see Fig.~\ref{fig:chandra_soft}). The eastern box is centred at 2:41:05.501, -8:15:21.65 (equatorial coordinates in J2000.0 epoch) with a size of $12.34 \times 8.07\arcsec$~$^2$, while the western box is  centred at 2:41:04.319, -8:15:21:30 with a size of $7.12 \times 8.30\arcsec$~$^2$.
 We underline that these areas exclude the central region where the nuclear emission dominates (see Fig. \ref{fig:chandra_soft}).
Given the low number of counts in these regions (346 and 262 counts in the east and west, respectively), we binned the spectra to have one count per bin and used Cash statistics for the analysis. The spectra were fitted over the $0.5-7$~keV energy range as there was no detection above 7~keV. We also extracted spectra in regions located to the north and south of the nucleus, but found that the spectra did not have sufficient signal for analysis.

\subsection{XMM-Newton observations}

The details of the four  {\it XMM-Newton} observations of NGC~1052 analysed in this work are summarised in Table~\ref{tab:obs}. There is also a more recent {\it XMM-Newton} observation available (from 2017 January 17), but we did not analyse this since the  source lies very close to a chip gap in the pn and overlaps with a bright column in MOS2.  
The four observations analysed are separated by seven months to five years, which is suitable for investigating possible spectral variability of the nucleus. From these observations we have used all data except those where the source falls on a chip gap, which turns out to be the case for the first MOS observation in 2001 and the MOS2 observation from 2009 January 14. All the observations were performed in full frame mode with the medium filter.

For the data reduction we processed the observation data files using {\texttt SAS}~v15.0.0 and the calibration files XMM-CCF-REL-349, following  standard procedures.\footnote{'Users Guide to the XMM-Newton Science Analysis System', Issue 15.0, 2019 (ESA: XMM-Newton SOC)}  All the observations were moderately affected by time periods with strong background flares, which we excluded from the analysis. These time periods correspond to $\sim30$ per cent of the total exposures (see Table \ref{tab:obs}). We also verified that the observations were not affected by pileup using the tool EPATPLOT. Source spectra were extracted from circular regions with radii of 55$\arcsec$ (following \citealt{evans2007} for bright sources observed with {\it Swift}, which has an angular resolution similar to {\it XMM-Newton}). The only exception to this was when the source fell close to a chip gap, in which case we reduced the extraction region to 30$\arcsec$. The background spectra were extracted from circular regions with $\sim80 \arcsec$ radii located near the source on the same detector chip.

The spectra from the MOS1 and MOS2 instruments were merged with a similar procedure as in \cite{falocco2014}. The merging procedure computes the sum of the exposures and the counts, while the backscale is the exposure-weighted  average of the backscales of the individual spectra.  The RMFs and ARFs  were similarly combined using   the {\texttt ftools} task {\texttt ADDRMF}. The resulting spectra were binned in order to have at least 25 counts per bin and to not over-sample the spectral resolution by more than a factor three.  The spectral analysis was performed over the $0.5 - 10$~keV energy range using $\chi^2$ statistics.

\subsection{Radio data}

We used radio observations from the Very Large Array (VLA) performed on 2002 January 26 (proposal code AC0622).   The data consist of a short observation with a total exposure time of 5 min on the source. The observations were performed in A configuration in the L band (central observing frequency 1.4~GHz and bandwidth 50~MHz). The beam size is $1.79\times 1.33"$.

\section{Data analysis and results}

We start the analysis by investigating the morphology of the galaxy and jet using {\it Chandra} and VLA images in Section~\ref{sec:images}. We then present the X-ray spectral analysis in Section~\ref{sec:spectra}, where we first characterise the {\it Chandra} spectrum of the extended emission and then analyse the {\it XMM-Newton} and {\it Chandra} spectra of the nucleus. We focus our analysis on time-averaged spectra, which allow us to investigate possible long-term variations. The light curves of the individual {\it XMM-Newton} observations (except the first short one) have already been studied by \cite{hernandez-garcia2014}, who showed that no significant variability was present during the observations.

\subsection{Imaging}
\label{sec:images}

\begin{figure*}
 \resizebox{!}{48mm}{\includegraphics[scale=1.2]{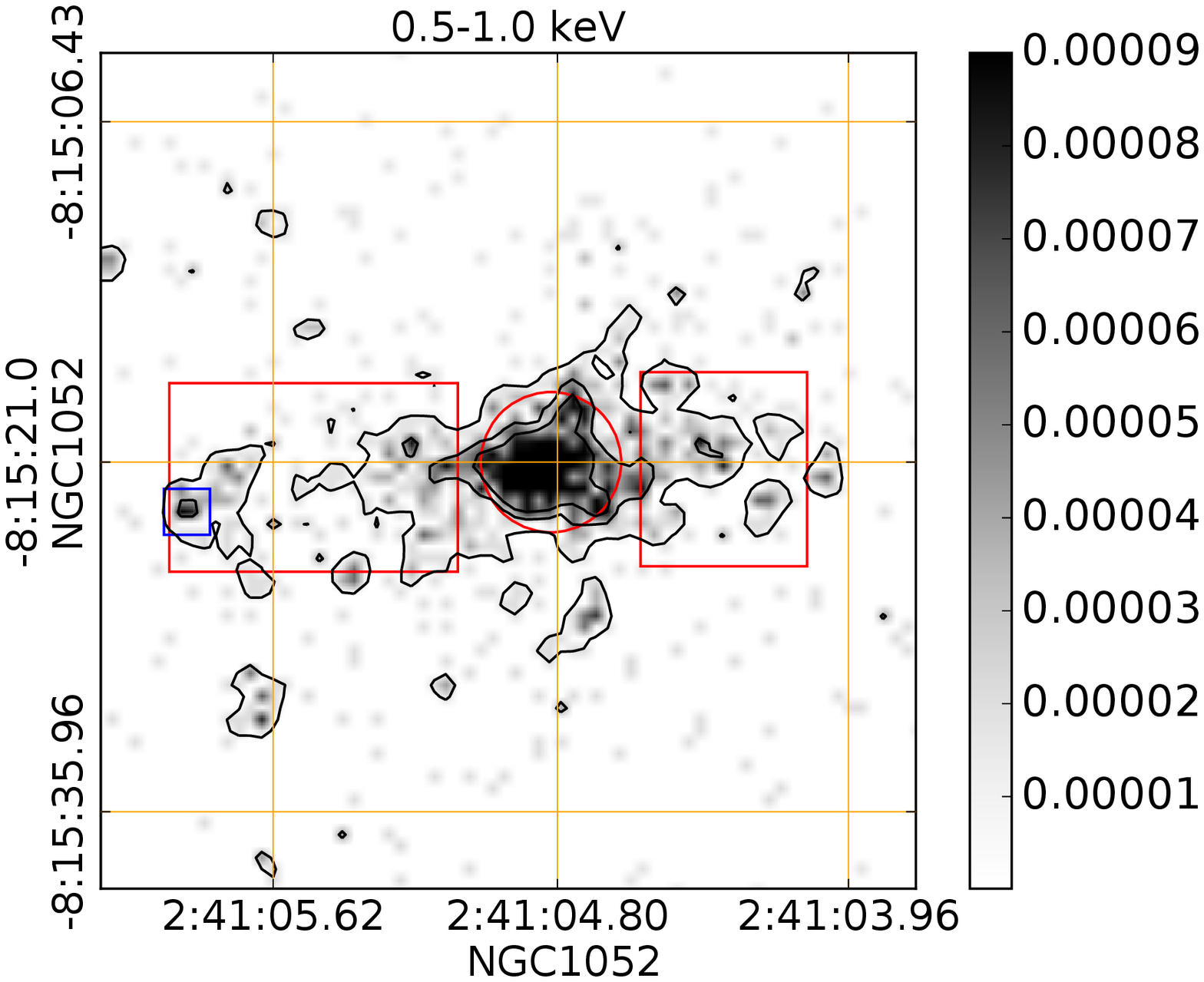}} 
\hspace{-0.4cm}
\resizebox{!}{48mm}{\includegraphics[scale=1.2]{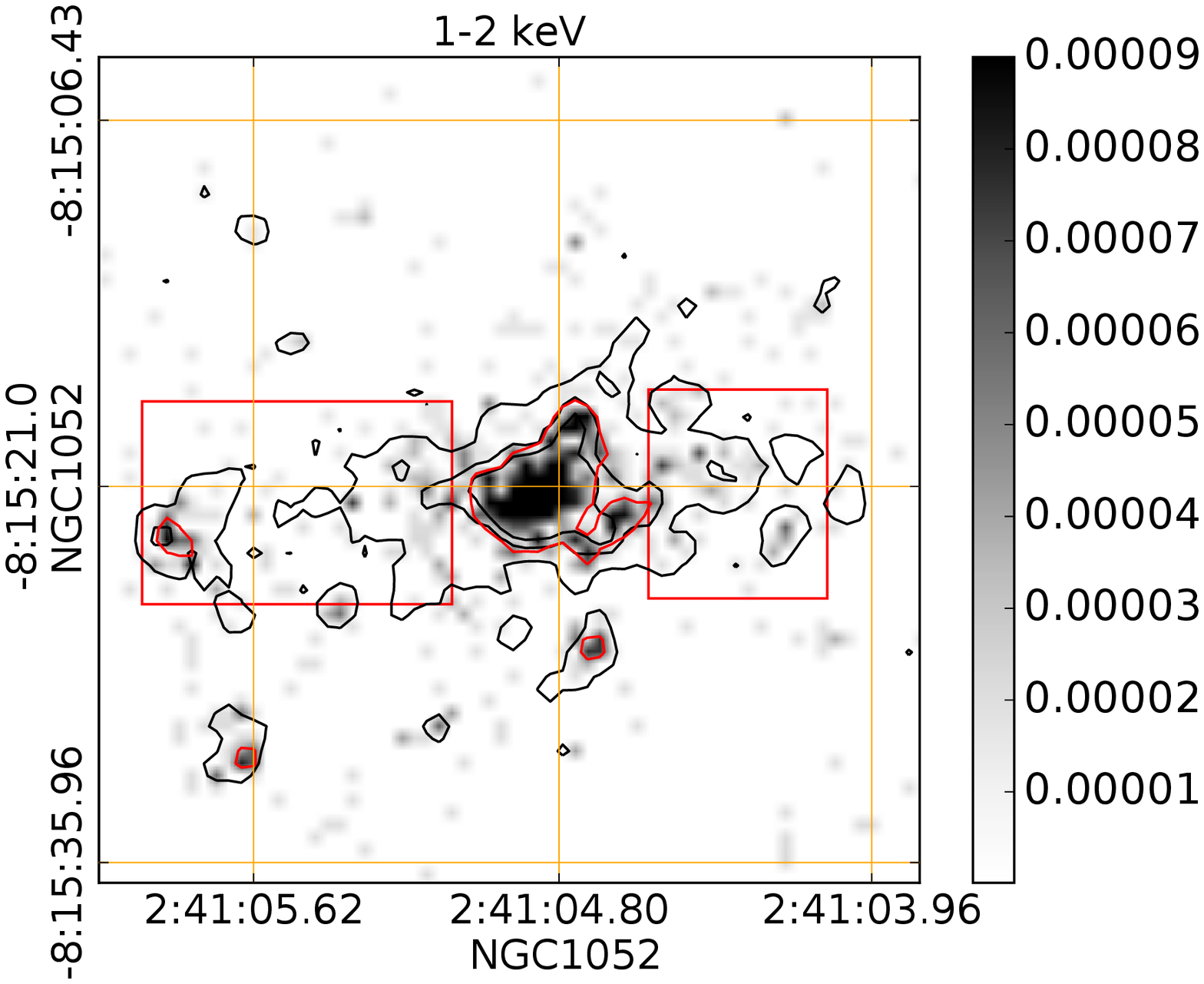}}
\hspace{-0.4cm}
 \resizebox{!}{48mm}{\includegraphics[scale=1.2]{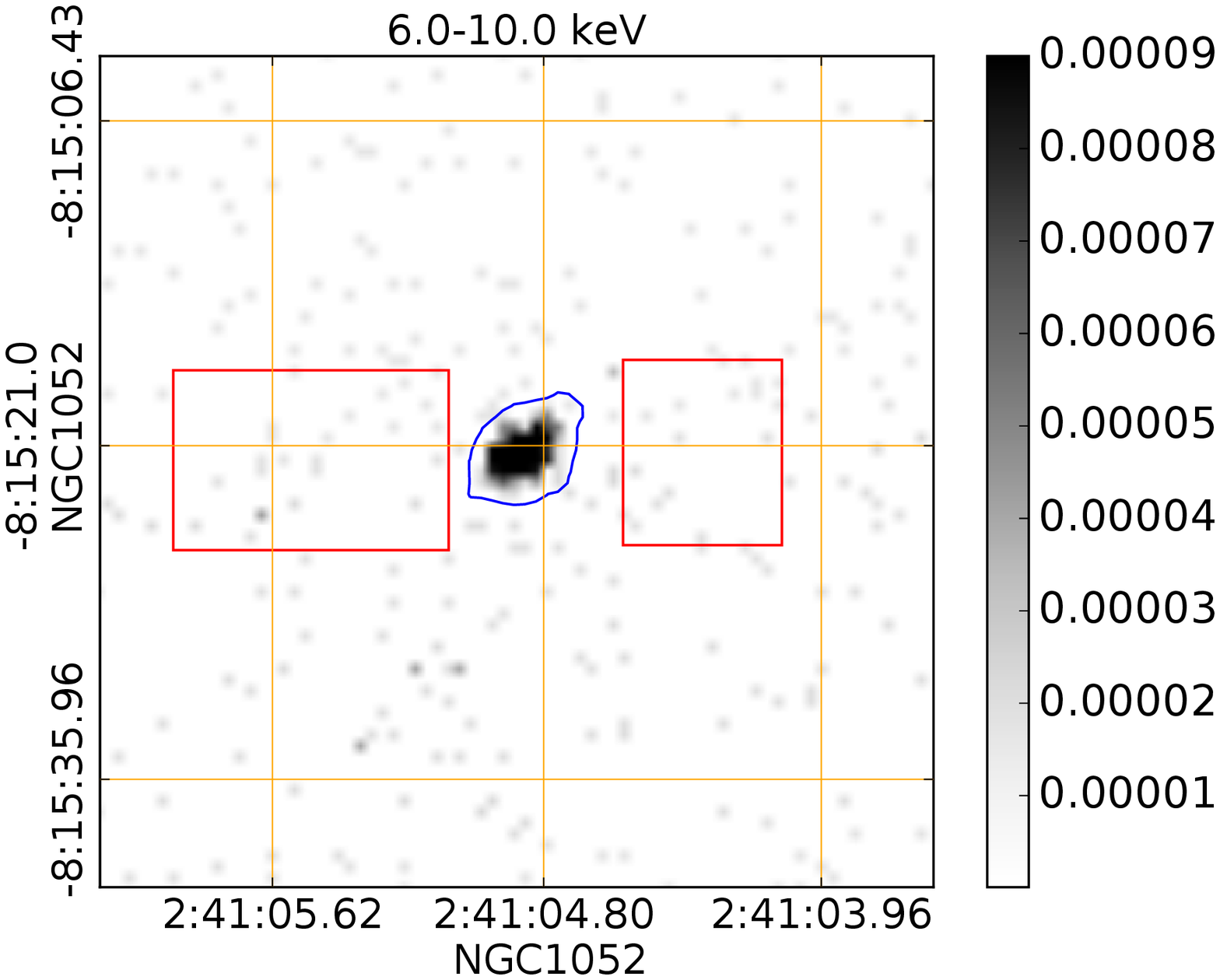}}
 \caption{
{\it Chandra}  images of NGC~1052 (units of cts~s$^{-1}$) covering the central $22\arcsec\times27\arcsec$ region ($\sim2.4\times2.9$ kpc). The orange grid shows units of $11.1\arcsec \times 13.3\arcsec$, 
where 11.1$\arcsec$ corresponds to 1.2~kpc at z=0.005. Left: Image in the $0.5-1$~keV energy range together with black contours corresponding to flux levels at  5, 20 and 35~$\sigma$ above the background. The red circle shows the outer radius of the spectral extraction region for the nucleus and the red rectangles show the spectral extraction regions used for the extended emission from the jet and galaxy. A hotspot associated with the eastern radio lobe is highlighted by a blue rectangle (the radio lobe can be seen in Fig.~\ref{fig:chandra_soft_radio}). Centre:  image in the $1-2$~keV energy range together with a red contour corresponding to 5$\sigma$. The contours for the 0.5-1 keV image in the left panel are also shown for reference. Right: Image in the $6-10$~keV energy band. The blue contour shows the 5$\sigma$ level from the $2-6$~keV image.}
\label{fig:chandra_soft}
  \end{figure*}

\begin{figure*}
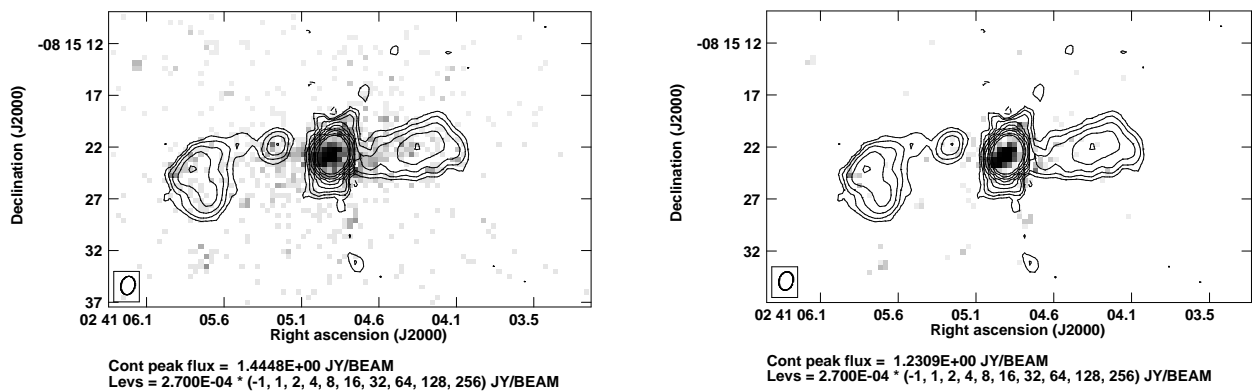

\rotatebox{270}{\resizebox{!}{80mm}{\includegraphics{selected_figures/RV1NGC1052_0.5-1.0.PS}}}
\hspace{0.5cm}
\rotatebox{270}{\resizebox{!}{80mm}{\includegraphics{selected_figures/RV1NGC1052_1.0-2.0.ps}}} 
\caption{Comparison of {\it Chandra} and VLA images of NGC~1052. The left and right panels show the {\it Chandra} images over $0.5-1$ and $1-2$~keV, respectively (see also Fig.~\ref{fig:chandra_soft}). The superposed contours are from the VLA 1.4~GHz observation from 2002. Contour levels are 3 $\sigma$ ( -1, 1, 2, 4, 8.... 256), where $\sigma$=0.09mJy/beam the image root mean square.}
\label{fig:chandra_soft_radio}
  \end{figure*}

We produced exposure-corrected images from {\it Chandra} using the  {\texttt FLUXIMAGE} tool in  {\texttt CIAO}, using monochromatic energies to calculate the exposure maps. The images were extracted in different energy bands in order to probe a variety of different emission components, including the AGN-like nucleus, the jet, as well as the galaxy, where X-ray emission may arise from the ISM, star forming regions, binaries etc. We used the energy bands $0.5-1$, $1-2$, $2-6$ and $6-10$~keV, with nominal energies of 0.9, 1.5, 3.8 and 8.0~keV, respectively.  The images are plotted in Fig.~\ref{fig:chandra_soft}, where the two highest energy bands are plotted together in the right panel. Comparing the images, we note that the emission becomes less extended as the energy increases, with the radiation from the nucleus dominating at the highest energies.  At energies below 1~keV, we see diffuse emission from the galaxy over an extended scale.  Between 1 and 2~keV, there is less extended diffuse emission as well as a small number of discrete sources. Only the nucleus is clearly detected in the images above 2~keV.

The X-ray emission at low energies extends $\sim 10\arcsec$ from the nucleus, with most of the emission being concentrated along the east-west direction.  The X-ray emission is more compact than the optical galaxy, which has a diameter of  $\sim100\arcsec$ according to NED\footnote{The NASA/IPAC Extragalactic Database (NED) is operated by the Jet Propulsion Laboratory, California Institute of Technology, under contract with the National Aeronautics and Space Administration}.   However, the X-ray emission is similar in scale to the radio jets.  This is seen in Fig.~\ref{fig:chandra_soft_radio}, where we compare the VLA 1.4~GHz observation from 2002 with the {\it Chandra} images in the $0.5-1$ and $1-2$~keV energy ranges. The VLA image shows large scale radio structure which is extended over 2.82 kpc.
 The radio jet has also been studied on pc-scales using VLBI observations (e.g., \citealt{kadler2004h,bazko2016,baczko2019}), but these small scales are not resolved with {\it Chandra}. 

 The VLA image at 1.4 GHz shows two compact high surface brightness regions, or two hotspots, on the eastern lobe. These have also been seen in a MERLIN observation from 1995 (\citealt{kadler2004}, labelled as H1 and A in their Fig.~3). On the western lobe, one main hotspot is seen, with some indication of an inner secondary one. The latter was seen more clearly in the MERLIN observation \citep{kadler2004}. In the X-ray images, a hotspot below 2~keV is seen along the outer edge of the eastern radio lobe.  This X-ray emission is likely due to shocks in the region where the extended jet interacts with the environment, as discussed in \citep{kadler2004}. On the same lobe,  there is also some X-ray emission below 1~keV at the position of the inner radio hotspot. However, it is not clear if this is associated with the jet since X-ray emission at a similar level is also seen in the nearby regions that do not overlap with the radio contours. The western radio lobe has a less clear counterpart in the X-ray band, but we note some low surface brightness emission coincident with the southern boundary of the lobe.

\subsection{Spectral analysis}
\label{sec:spectra}

We performed the spectral analysis using \texttt{ XSPEC} version 12.9. All fits described below include Galactic absorption modelled with \texttt{tbabs}. The Galactic H column density in the direction of NGC~1052 is $3.1 \times10^{20}~\rm{cm^{-2}}$, determined using the \texttt{NHTOT} tool \footnote{http://www.swift.ac.uk/analysis/nhtot/docs.php}, which takes into account the contribution from molecular hydrogen. We first analyse the {\it Chandra} spectra of the extended jet plus galaxy regions, in order to use the best-fitting model for the analysis of the {\it XMM-Newton} spectra, where the galaxy is unresolved.

The {\it XMM-Newton} EPIC pn and MOS spectra were fitted simultaneously, with all parameters tied except for a cross-normalisation factor. We finally disuss the {\it Chandra} spectrum of the nucleus. This has relatively low signal-to-noise, and the analysis is therefore guided by the results from the {\it XMM-Newton} spectra. 
We quote uncertainties on fit parameters at 90~per~cent significance for one interesting parameter ($\Delta \chi^2 = 2.7$).

\subsubsection{Chandra spectra of the extended emission}
\begin{figure}
    \includegraphics[width=8cm,height=9cm,angle=-90]{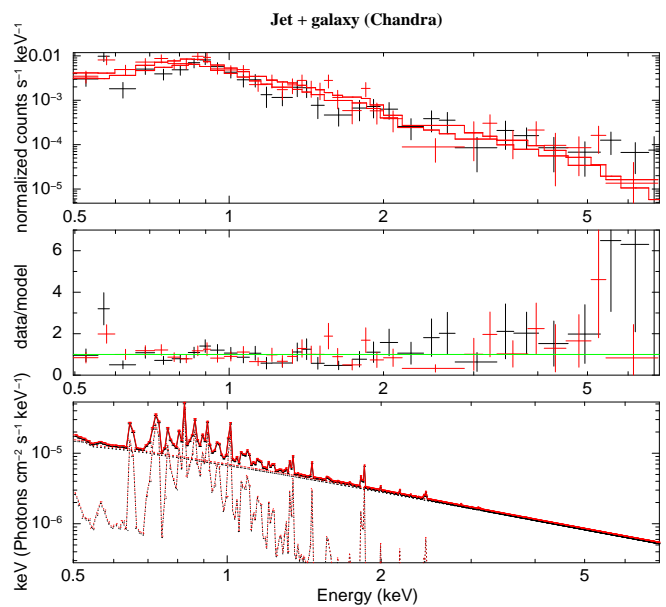}
    \caption{{\it Chandra} spectra of the extended emission from the galaxy plus jet fitted with the  \texttt{pow+ mekal}  model. The red and black spectra were extracted from the regions east and west of the nucleus, respectively (see Fig.~\ref{fig:chandra_soft}). The spectra have been binned for visual purposes only (to have 4~$\sigma$ significance in each bin). Top panel: spectra with the best-fit model, central panel: ratio of data/model, bottom panel: model components.}
  \label{fig:chandra}
  \end{figure}

The spectra of the extended emission extracted from the two rectangular regions in Fig.~\ref{fig:chandra_soft} show very similar features. We therefore fitted them together with all parameters tied except for a cross nomalisation constant. Due to the limited statistics of the spectra we only considered simple models in the form of a power law (expected from jet emission) and a thermal \texttt{mekal} component. The latter may arise from the ISM, shocks or starburst regions. The results from the fits are reported in Table~\ref{tab:chandra_jgalfits}. We found that a single component could not describe the spectra, whereas a \texttt{pow+mekal} and \texttt{mekal+mekal} provided acceptable fits. We adopt the former as our preferred model due to slightly better cstat and the high temperature of the second \texttt{mekal} component in the latter model, which lacks a clear physical interpretation.   We also tested adding intrinsic absorption to the models, but found that it was not required.
The spectra and best-fitting model are shown in Fig.~\ref{fig:chandra}.
We refer to our preferred model for the jet plus galaxy region as JGAL hereafter.
        
\renewcommand{\arraystretch}{1.5}
\begin{table}
\centering
\caption{Results from fits to the {\it Chandra} spectra of the extended emission from the jet and galaxy. The fluxes are unabsorbed. }
\label{tab:chandra_jgalfits}
\begin{tabular}{lc} 
 \hline
 Parameter     &  Value  \\ 
\hline
\texttt{tbabs*mekal}   &    \\
$T$ [keV]    &     $0.79^{+0.03}_{-0.04}$  \\
cstat/d.o.f.   &  477.8/248   \\
\hline

\texttt{tbabs*(mekal+mekal) }   &      \\
  $T_{\rm h}$ [keV]    &     $5.3^{+5.7}_{-1.9}$      \\
$T_{\rm c}$  [keV]   &     $0.32^{+0.03}_{-0.03}$      \\
 cstat/d.o.f.   &  269.8/246  \\
 \hline 
 
\texttt{tbabs*pow}    &     \\
$\Gamma$     &     $2.7^{+0.1}_{-0.1}$   \\
cstat/d.o.f.   &  292.4/248   \\
 \hline
 
\texttt{tbabs*(pow+mekal) }   &     \\
$\Gamma$   &         $2.3^{+0.2}_{-0.1}$     \\
 $T$ [keV]   &                 $0.6^{+0.1}_{-0.1}$     \\
cstat/d.o.f.    &  258.4/246   \\
   
$F_{\rm tot,0.5-7}$   [erg$~\rm{cm^{-2}~s^{-1}}$]     &   $6.8\pm 1.1\times10^{-14}$   \\
$L_{\rm tot,0.5-7} $  [erg$~\rm{s^{-1}}$]     &    $3.5\pm 0.6\times10^{39}$  \\
$L_{\rm th,0.5-7}$   [erg$~\rm{s^{-1}}$]     &   $ 8.6\pm 2.6 \times10^{38}$   \\

    $F_{\rm{tot,0.5-2}}$   [$\rm{erg~s^{-1}~cm^{-2}}$]     &  $4.9\pm0.9 \times10^{-14}$    \\
 $L_{\rm{tot,0.5-2}}$   [$\rm{erg~s^{-1}}$]    &  $2.4\pm0.5 \times10^{39}$    \\
   
\hline
\end{tabular}
\end{table}

 \begin{figure*}
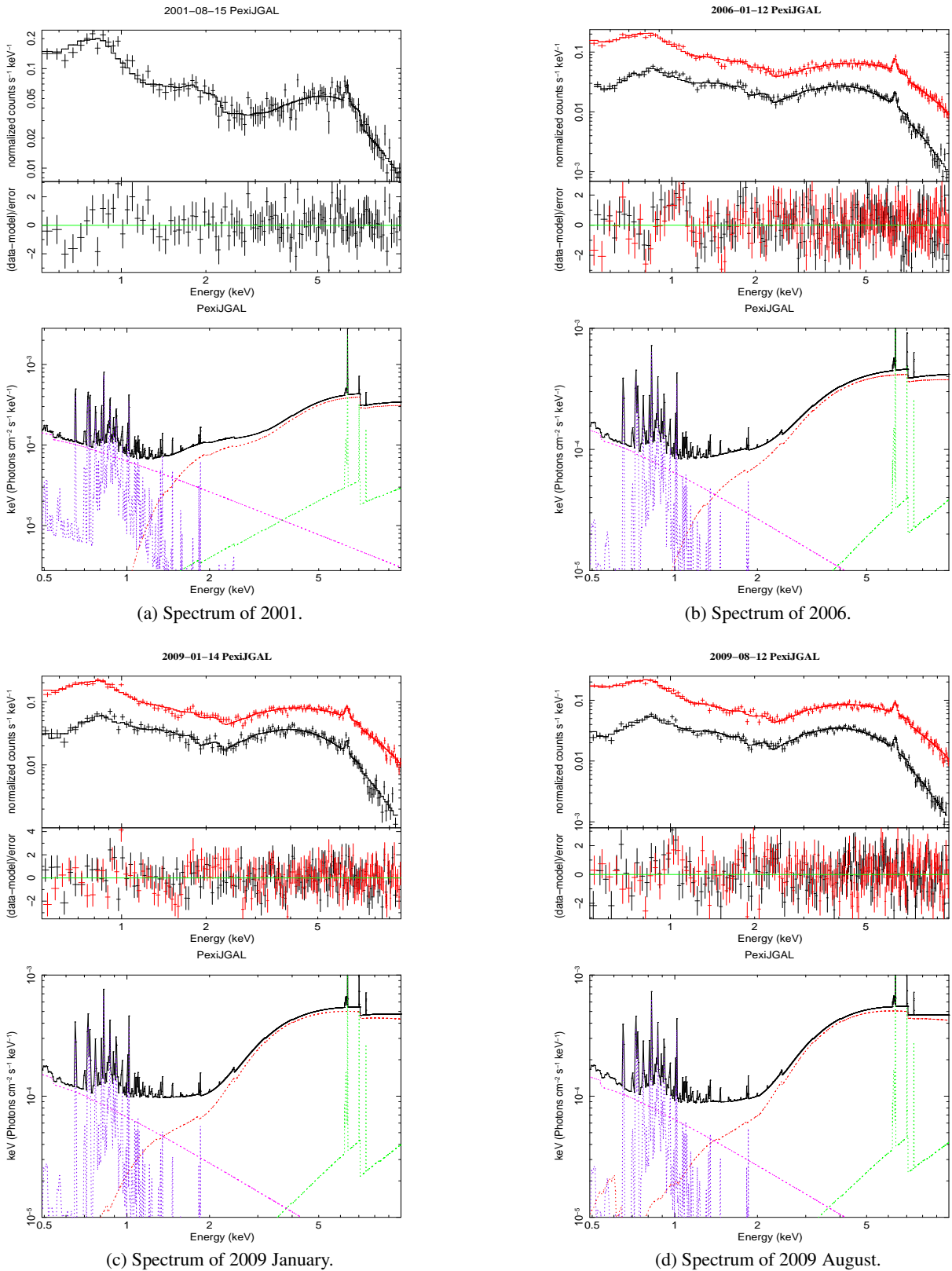

\subfloat[Spectrum of 2001.]{
  \begin{minipage}[b]{8cm}
    \includegraphics[width=5cm,height=7cm,angle=-90]{selected_figures/0093630101_PN_grp_pow_pexm_ion_fitted_05_10_simple_jet.ps}
     \includegraphics[width=5cm,height=7cm,angle=-90]{selected_figures/0093630101_PN_grp_pow_pexm_ion_fitted_05_10_simple_jet_emo.ps}
  \label{fig:xmm1}
\end{minipage}}
   \hspace{1cm}
 \subfloat[Spectrum of 2006.]{   
\begin{minipage}[b]{8cm}
    \includegraphics[width=5cm,height=7cm,angle=-90]{selected_figures/0306230101_grp_pow_pexm_ion_fitted_05_10_refitted_simple_jet.ps}
    \includegraphics[width=5cm,height=7cm,angle=-90]{selected_figures/0306230101_grp_pow_pexm_ion_fitted_05_10_refitted_simple_jet_one_spec_em.ps}
  \label{fig:xmm2}
\end{minipage}}
 \\
 \subfloat[Spectrum of 2009 January.]{ 
\begin{minipage}[b]{8cm}
    \includegraphics[width=5cm,height=7cm,angle=-90]{selected_figures/0553300301_grp_pow_pexm_ion_fitted_05_10_refitted_simple_jet.ps}
    \includegraphics[width=5cm,height=7cm,angle=-90]{selected_figures/0553300301_grp_pow_pexm_ion_fitted_05_10_refitted_simple_jet_one_spec_em.ps}
    \label{fig:xmm3}
\end{minipage}}
 \hspace{1cm}
 \subfloat[Spectrum of 2009 August.]{ 
\begin{minipage}[b]{8cm}
    \includegraphics[width=5cm,height=7cm,angle=-90]{selected_figures/0553300401_grp_pow_pexm_ion_fitted_05_10_refitted_simple_jet.ps}
    \includegraphics[width=5cm,height=7cm,angle=-90]{selected_figures/0553300401_grp_pow_pexm_ion_fitted_05_10_refitted_simple_jet_one_spec_em.ps}
\end{minipage}}
   
 \caption{{\it XMM-Newton} spectra fitted with the PexiJGAL model (black line), together with the fit residuals and the individual model components. 
The red line is the  absorbed power law from the nucleus and the green line is the \texttt{pexmon} model. The blue and magenta lines are the jet plus galaxy model}
\label{fig:xmm_allplots}
  \end{figure*}

\subsubsection{XMM-Newton spectra}
The {\it XMM-Newton} spectra of NGC~1052 include emission from the nucleus, extended jet and surrounding galaxy. To account for the latter components we included the {\it Chandra} JGAL model described above in all fits to {\it XMM-Newton} spectra. We kept all parameters of JGAL fixed at their best-fit values (from Sect. 3.2.1) except for the normalisation (but keeping the same relative normalisation between the two components as in {\it Chandra}). We started by modelling the emission from the nucleus as a partially absorbed power law and a narrow Fe line at 6.4~keV, which is the simplest possible model motivated by previous studies of the source \citep{guainazzi1999,guainazzi2000,brenneman2009}. The absorption is assumed to be caused by the molecular torus (e.g., \citealt{baczko2019}). In  \texttt{XSPEC} the model is  \texttt{tbabs*(zpcfabs*pow+zgauss+mekal+pow)}. We will refer to it as PG-JGAL from here on. The fit results are  reported in Table~\ref{tab:xmm_fits_norefl} in the Appendix. 
The fitting procedure starts by characterising the continuum between $0.5-5.5$ and $7-10$~keV before adding the Gaussian line and fitting the whole energy interval. When adding the line, its energy was fixed at 6.4~keV in order to represent the K$\alpha$ line from neutral iron. Its width was fixed at 0.09~keV to be intrinsically narrow and broadened only by the spectral resolution of the instruments. Allowing the energy and width of the line to vary did not significantly change the results, confirming a narrow line from neutral iron. The $\chi^2$ values obtained when fitting the model without the line are reported in the first row of Table~\ref{tab:xmm_fits_norefl} (where the model without the line is referred to as P-JGAL). A comparison with the fit statistic for the PG-JGAL model in the same table shows that the iron line is highly significant, improving the $\chi^2$ by $\sim 100-140$ for one degree of freedom (d.o.f) for the three observations with the longest exposures.

Since the iron line is highly significant, we introduced a complete model which considers Compton reflection and fluorescence in a self consistent way, considering that Compton reflection is expected together with fluorescence.
In line with this, \cite{osorio2019} found evidence for Compton reflection in {\it Nustar} and {\it Suzaku} observations of the source. To account for this reflection, we substituted the gaussian line in our model with a \texttt{pexmon} component \citep{nandra2007}. We refer to this model as PexJGAL. \texttt{Pexmon} includes fluorescent lines of Fe K$\alpha$, FeK$\beta$, and Ni K$\alpha$, the Fe K$\alpha$ Compton shoulder, as well as Compton reflection. We allowed only the normalisation of the \texttt{pexmon} component to vary in the fits. The inclination of the source is known to be high \citep{baczko2019} and we fixed it at 85$^\circ$, noting that its exact value does not affect the conclusions. We further tied the photon index to the value of the primary power law and kept the cutoff energy and abundances fixed at their default values. 
This model does not produce a satisfactory $\chi^2$ (see Table \ref{tab:xmm_fits}) and leaves residuals especially in the soft band below 3~keV. For this reason we included an additional absorption component, as also done in the previous {\it Suzaku} observation \citep{brenneman2009}.
 The overall model in \texttt{xspec} was: \texttt{tbabs(zpcfabs*absori*pow + pexmon + mekal + pow)}, hereafter referred to as PexiJGAL. 
 
 The model PexiJGAL produces acceptable fits to all spectra, as can be seen in Table ~\ref{tab:xmm_fits}. The reflection fraction\footnote{The reflection fraction is derived as the ratio between the fluxes of the reflected radiation to the total AGN radiation between 2 and 10 keV} ranges between 3 and 6 \%.
Fig. \ref{fig:xmm_allplots} shows the spectra fitted with this model, along with the individual model components, and Table \ref{tab:xmm_fits} reports the fits parameters. 
The value of the ionisation parameter of the \texttt{absori} component is consistent with null ionisation ($\xi < 2~{\rm erg~cm~s^{-1}}$, see Table~\ref{tab:xmm_fits}), showing that the second absorber is consistent with being neutral. Indeed, replacing \texttt{absori} with a cold absorption model (\texttt{pcfabs} with 100~per~cent covering fraction) does not significantly change the fit statistic or the other model parameters. Independently of the ionisation degree, the second absorber significantly improves the fit for all {\it XMM-Newton} spectra, resulting in  $\Delta \chi^2$ of $>40$ for 2~d.o.f., (Table~\ref{tab:xmm_fits}).
We discuss these results in Sect. 4.3. 

Although the PexiJGAL model works well for all spectra, there are still some discrepancies between the observed spectra and the model in the soft band.
This is particularly evident in the {\it XMM-Newton}  spectra of 2006 and of 2009-08-15. We thus tested some modifications to the PexiJGAL model applied to the spectrum of 2009-08-15, which has the best spectral quality.
As a first test we allowed the abundances of the mekal component in the JGAL model to vary. We found that the abundances were poorly constrained and that the improvement in the fit was not significant.
 We then performed a test leaving the abundances of the JGAL model fixed as before, but adding an extra \texttt{mekal} component. We found that a collisional plasma at a temperature $1.3\pm0.2$ keV significantly improves the fit  ($\Delta\chi^2=32$ for 2 d.o.f). There was no significant change in the other model parameters, showing that the continuum characterisation with PexiJGAL is robust.
 The additional plasma contributes at most 9 \% to the total flux (unabsorbed flux from the AGN plus JGAL component) between 0.5 and 2 keV, which is smaller than uncertainties on the total flux itself ($\sim$13\% in this band). Considering that the test was performed in the best observation, we conclude that the addition of this component would only have a minor effect in the fits to the other spectra.

 {\it Constraints on relativistic reflection}
 
  We have tested relativistic reflection components in addition to the PexiJGAL model in order to investigate if any reflection features from the inner accretion disk close to the SMBH are present in the {\it XMM-Newton} spectra.
  We performed the tests only with the first and third observations (OBSIDs 0093630101 and 0553300301). This is because the first spectrum shows an indication of an excess on the red side of the narrow  line (see Fig.~\ref{fig:xmm_allplots}, panel a), while the third one has a good spectral quality (see Fig.~\ref{fig:xmm_allplots} panel c).

We started by adding only a \texttt{laor} model, which represents the line emitted from a disk around a Kerr black hole \citep{laor1991}. We found that this model did not improve the fit and that the free parameters could not be constrained, showing that no prominent broad line is present. To still obtain some constraints on the line we proceeded by fixing most of the parameters, using information obtained from other studies of the source. Radio observations have shown that the inclination, $i$, is high \citep{bazko2016,baczko2019}, although \cite{baczko2019} finds that there is no value of $i$  that can explain all observations, suggesting that the jets are asymmetric. We therefore performed the fits with $i$ fixed at different values in the range $65^{\circ} \leq i \leq 90^{\circ}$, finding that it did not affect the conclusions. We initially also fixed the inner radius of the disk at $r_{\rm{in}}=26~r_{\rm{g}}$,\footnote{$r_{\rm{g}}$ is the gravitational radius $= \frac{GM}{c^2}$} which is a lower limit estimated by comparing the accretion power of the disk with the spectral energy distribution from  sub-arcsec resolution observations \citep{reb2018}. Finally, we fixed the the outer radius of the disk ($r_{\rm{out}}$) and the emissivity index ($q$) at their default values of $r_{\rm{out}} =400~r_{\rm{g}}$ and $q=3$. This leaves the normalisation as the only free parameter. This model did not significantly improve the fit, resulting in a very similar $\chi^2$ and an equivalent width (EW) of the line consistent with zero.   Leaving the inner radius free to vary did not change the results, as well as fixing it to its lower value 1.24 $r_g$.

We also performed fits where we replaced \texttt{laor} with the \texttt{relxill} model \citep{dauser2014}, which derives the full reflection spectrum from the disc, including line emission and Compton scattering. We fixed $i$, $q$, $r_{\rm in}$ and $r_{\rm{out}}$ at the same values as for the \texttt{laor} model.
In addition, the photon index of the primary radiation was tied to the value of the power-law model, the ionisation was set to zero, the iron abundance to 1 (in units of solar abundances) and the high-energy cutoff of the primary spectrum was fixed to 300~keV.
Only the normalisation was thus left free to vary in \texttt{relxill}. 
This test produced a marginal improvement of the $\chi^2$ ($\Delta\chi^2$ of 3 in the first observation and 2 in the third one for 1 d.o.f. less than the fit with PexiJGAL). Leaving $r_{\rm{in}}$ free did not result in any significant improvements.  Its best fit value is $ 14~r_{\rm{g}}$ in both observations with upper limits of 370 and 400$~r_{\rm{g}}$ in the first and third observation, respectively.
Since the model PexiJGAL includes Compton reflection from distant material, we have also tested for the presence of a relativistic line after excluding the \texttt{pexmon} component. This allows us to investigate if part of the broad line had been modelled as Compton reflection. We thus added the laor model (with the same constraints as above) to the model referred to as PGI-JGAL in Table \ref{tab:xmm_fits_norefl} in the Appendix.
We found that the inner radius is constrained to $< 100~r_{\rm{g}}$ in the first observation and to $< 90~r_{\rm{g}}$ in the third one, with a $\chi^2$ similar to the fit  without the relativistic line. This confirms that the relativistic line is not significantly required by the data.

  \subsubsection{Chandra spectrum of the nucleus}
  
  We finally analysed the {\it Chandra} spectrum of the nucleus. This spectrum has significantly lower signal than the {\it XMM-Newton} spectra since the brightest part of the PSF was excluded in order to avoid pileup. The {\it Chandra} spectrum also differs from the {\it XMM-Newton} spectra in that the contribution from the jet+galaxy is expected to be much smaller due to the superior spatial resolution. The results from the fits to the {\it Chandra} spectrum are summarised in Table~\ref{tab:chandra_results}. We started by fitting a model composed of an absorbed power law and a Gaussian, equivalent to our initial model for the {\it XMM-Newton} spectra without the jet plus galaxy model.
  
\begin{figure}
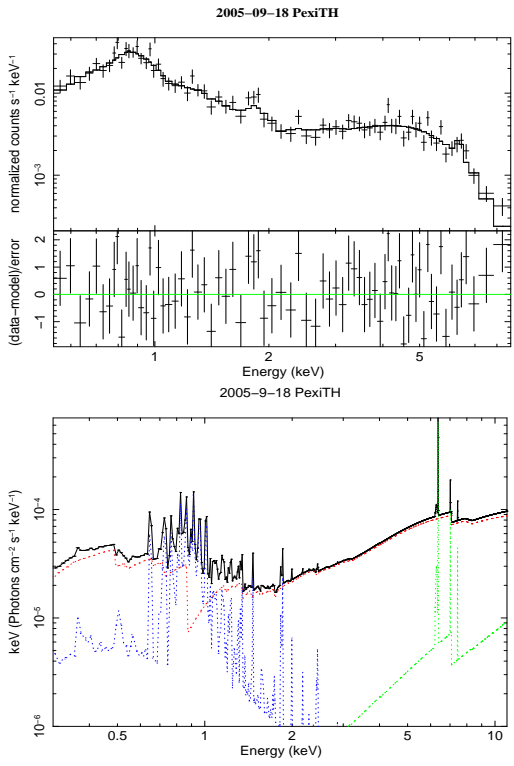

    \includegraphics[width=5cm,height=7cm,angle=-90]{selected_figures/chandra_grp_pow_pexm_ion_mekal_fitted_05_10.ps}
    \includegraphics[width=5cm,height=7cm,angle=-90]{selected_figures/chandra_grp_pow_pexm_ion_mekal_fitted_05_10_emo.ps}
    \caption{The {\it Chandra} spectrum of the nucleus from 2005 fitted with the PexiTH model (black line). The model components are shown in the lower panel. 
      The red line is the  absorbed power law from the nucleus, the green line is the \texttt{pexmon} model, and the blue line is the \texttt{mekal} component from the galaxy.}
  \label{fig:chandra_nucleus_fit}
  \end{figure}

    Since we found that the narrow iron line is significant, we substituted the gaussian with the self-consistent \texttt{pexmon} model as was done for the {\it XMM-Newton} spectra. Since there were
    residuals in the soft X-rays, we added a thermal model to represent a contribution from the galaxy.
 The\texttt{ xspec} model was \texttt{tbabs(zpcfabs*powerlaw + pexmon + mekal)}, hereafter called PexTH. The fit results are presented in Table~\ref{tab:chandra_results}. 
To take into account some remaining residuals in the soft X-rays and to use a model consistent with the one used for {\it XMM-Newton}, we finally added an ionised absorber. The xspec model was \texttt{tbabs(zpcfabs*absori*powerlaw + pexmon + mekal)}, hereafter called PexiTH. The fits results are reported in Table~\ref{tab:chandra_results} and the spectrum with the best fitting model is shown in Fig.~\ref{fig:chandra_nucleus_fit}.

The primary radiation of the Chandra spectrum of the nucleus is rather flat compared to typical values observed for standard AGN. However, it is consistent with the $\Gamma$ of the second {\it XMM-Newton} spectrum, which is the closest in time (observed four months later) within the uncertainties. The column densities of the low and high density absorbers are poorly constrained, with best-fit values similar to the first {\it XMM-Newton} observation (observed four years earlier), but higher than the other observations with better spectral quality. The flux of the AGN component in Table~\ref{tab:chandra_results} has been corrected for the fact that the {\it Chandra} extraction region is expected to exclude 80 per cent of the flux from the PSF. The corrected flux is consistent with the AGN fluxes obtained from the {\it XMM-Newton} spectra.

The fact that the {\it Chandra} spectrum of the nucleus requires a thermal component shows that there is a significant contribution from the galaxy also in the nuclear region. The temperature of this component is consistent with that found in the fits to the spectra from the galaxy plus jet (cf. Tables  \ref{tab:chandra_jgalfits} and \ref{tab:chandra_results}). The total 0.5-2~keV flux from the emission components associated with the galaxy and the extended jet  in the {\it Chandra} spectra is $8.6 \times 10^{-14}~\rm{erg~cm^{-2}~s^{-1}}$ ($3.7 \times 10^{-14}~\rm{erg~cm^{-2}~s^{-1}}$ and  $4.9 \times10^{-14}~\rm{erg~cm^{-2}~s^{-1}}$ from the nuclear region and boxes, respectively, see Tables~\ref{tab:chandra_results} and \ref{tab:chandra_jgalfits}). This is somewhat lower than the $2.3 \times 10^{-13}~\rm{erg~cm^{-2}~s^{-1}}$ found in all four {\it XMM-Newton} observations (Table~\ref{tab:xmm_fits}), which is reasonable considering that the {\it XMM-Newton} spectra also include the centre of the nucleus (which was excluded for the {\it Chandra} spectrum of the jet plus galaxy squared regions) as well as the galactic contribution outside the squared regions (see Fig.~\ref{fig:chandra_soft}).

\section{Discussion}

In this Section we discuss the results considering several aspects raised from the imaging and spectral analysis. We start by discussing the star formation activity of the galaxy and its expected contribution to the X-ray emission in Section~\ref{sec:disc-star}, and then discuss the properties of the jets and the origin of the extended X-ray emission from the galaxy in Section~\ref{sec:disc-jet}. We finally discuss the nature of the nucleus in Section~\ref{sec:disc-core}, and compare with previous results in the literature in Sect. 4.4.

\subsection{Star formation}
\label{sec:disc-star}

Since part of the X-ray emission studied in this work is of galactic origin, we  start by exploring the star formation activity of the galaxy. The Star Formation Rate (SFR) of the galaxy can be constrained from the FIR luminosity, which was calculated from the fluxes at 60 and 100~$\mu$m (\citealt{ranalli2003} and references therein). The FIR fluxes are 0.9 Jy at 60~$\mu$m \citep{neugebauer1984} and 1.58 Jy at 100~$\mu$m (reported in NED as a private communication by Knapp), which correspond to a FIR luminosity of $2\times10^{42}~\rm{erg~s^{-1}}$  and a SFR of 0.09~$\rm{M_\odot~yr^{-1}}$. However, this SFR is an upper limit since the FIR luminosity is dominated by the AGN
  \citep{tang2009,asmus2014,fernandez-ontiveros2019}.  We also computed an upper limit to the specific SFR (sSFR)\footnote{The specific star formation rate is the ratio between the SFR and the stellar mass the galaxy} to have a measurement of the efficiency of star formation processes. Given the stellar mass of the galaxy of $10^{11}~\rm{M_\odot}$ \citep{fernandez-ontiveros2010}, the sSFR is extremely low, $9\times10^{-4}~\rm{Gyr^{-1}}$. Thus, the star formation is a rather inefficient process in NGC~1052. For comparison, in the sample of X-ray selected AGN of \cite{rovilos2012}, the lowest redshift sub-sample ($z <1.12$) has sSFRs spanning a broad range of values down to $0.1~\rm{ Gyr^{-1}}$ for the lowest luminosity sources with L$_{2-10~\rm keV} \sim4\times10^{41}~\rm{erg~s^{-1}}$.  Although the X-ray luminosity of NGC~1052 is within the low end of this sample,  its sSFR is well outside. In fact, the  sSFR of NGC~1052 classifies it as a 'red and dead' galaxy \citep{fontana2009}, as already noted in \cite{fernandez-ontiveros2010}. These galaxies have rather inefficient star formation, typically with  sSFR~$ <10^{-3}~\rm{Gyr^{-1}}$, characterised by long time-scales.

The X-ray luminosity is correlated with the FIR luminosity in highly star forming galaxies, while AGN and LINER powered by AGN  do not follow the starburst correlation due to their excess X-ray emission \citep{ranalli2003}. From the spectral fitting, the estimated $2-10$~keV AGN luminosity is $\sim 3-4 \times 10^{41}~\rm{erg~s^{-1}}$, Table~\ref{tab:xmm_fits}), significantly higher than expected for a starburst source. Indeed, from the FIR luminosity of $2 \times 10^{42}~\rm{erg~s^{-1}}$, the upper limit on the hard X-ray luminosity expected
from star formation activity is L$_{\rm 2-10~keV} = 3.8\times10^{38}~\rm{erg~s^{-1}}$ (from  the relation in \citealt{ranalli2003}). The jet plus galaxy component of the best-fit model to the {\it XMM-Newton} spectra has a luminosity of 4.5 $\times10^{39}\rm{erg~s^{-1}}$ between 2 and 10 keV (see Table~\ref{tab:xmm_fits}), which is much higher due to contributions different from pure star formation.

In the soft X-ray band between 0.5 and 2~keV,
the upper limit on the luminosity expected from pure star formation activity is $6.3\times10^{38}~\rm{erg~s^{-1}}$ \citep{ranalli2003}, while the  luminosity of the thermal component in the same band from our fits to the {\it XMM-Newton} spectra is $\sim 2.7-3 \times 10^{39}~\rm{erg~s^{-1}}$ (the thermal component contributes $\sim30$\% to the jet plus galaxy emission  between 0.5 and 2 keV reported in Table ~\ref{tab:xmm_fits}). This shows that processes other than pure star formation make a dominant contribution to the thermal soft X-ray emission from the galaxy, as discussed below.

\renewcommand{\arraystretch}{1.5}
\begin{table*}
  \centering
  \caption{Results of fits to the {\it XMM-Newton} spectra between 0.5 and 10~keV. The models are: PexJGAL:  \texttt{tbabs*(zpcfabs*pow+pexmon+mekal+pow)}. The last two \texttt{mekal+pow} components in the model represent the extended emission from the jet and galaxy and have parameters fixed to the model found for the jet and galaxy region from Chandra data (see Table~\ref{tab:chandra_jgalfits} and text for details).
    PexiJGAL:  \texttt{tbabs*(zpcfabs*absori*pow+pexmon+mekal+pow)}.
    $F_{\rm{AGN,0.5-10}}$ is the sum between the primary radiation and the reflected component. All fluxes and luminosities are unabsorbed.  N$_{\rm H,high}$ and  N$_{\rm H,low}$  are the column densities of the \texttt{zpcfabs} and \texttt{absori} models, respectively,  and $\xi$ is ionisation parameter of \texttt{absori}  (defined as $L/n~R^2$ where $L$ is the luminosity, $n$ the number density and $R$ the distance to the ionising source).  The luminosities are derived from the fluxes at a distance from NGC1052 of 19.1 Mpc (luminosity distance from NED). \emph{Refl} is t he reflection fraction, derived as the ratio between the fluxes of
the reflected radiation to the total AGN radiation between 2 and 10 keV. }
	\label{tab:xmm_fits}
    \begin{tabular}{l|r|r|r|r|r}\\

  PexJGAL Parameter    &   Unit &     0093630101  &  0306230101   &   0553300301 & 0553300401   \\
 \hline

  $N_{\rm{H}}$    & [$10^{22}~\rm{cm^{-2}}$]
  & $10.4^{+2.0}_{-1.7} $     &  $ 7.5^{+0.5}_{-0.4}$            &  $7.8^{+0.5}_{-0.5}$  &  $8.4\pm0.4$   \\   
    
    Cov-frac     &   & $0.85^{+0.02}_{-0.02} $   &  $0.88^{+0.01}_{-0.01}$  &  $0.92^{+0.01}_{-0.01}$ &  $0.94^{+0.01}_{-0.01}$ 
       \\  

       $\Gamma$ &   &  $<1.2 $   &  $<1.2$  &  $1.3^{+0.1}_{-0.1}$ &  $1.4^{+0.1}_{-0.1}$
       \\
       
       $\chi^2$/d.o.f.   &   & 168.0/118 &  432.5/309  & 409.5/298 & 426.7/317  \\
       
       \hline
       
  PexiJGAL Parameter    &    &       &     &    &    \\

      $N_{\rm{H,high}}$    & [$10^{22}~\rm{cm^{-2}}$]  & $25^{+6}_{-6} $   
    &  $ 10^{+1}_{-1}$            &  $8.7^{+0.7}_{-0.7}$  &  $9.1\pm0.5$   \\   
    
    Cov-frac     &   & $0.87^{+0.05}_{-0.05} $   &  $0.85^{+0.02}_{-0.02}$  &  $0.92^{+0.01}_{-0.02}$ &  $0.92^{+0.01}_{-0.01}$ 
       \\  

    $\Gamma$ &   & $1.82^{+0.30}_{-0.30} $   &  $1.34^{+0.09}_{-0.09}$  &  $1.45^{+0.09}_{-0.09}$ &  $1.53^{+0.08}_{-0.08}$    \\

     $N_{\rm{H,low}}$ & [$10^{22}~\rm{cm^{-2}}$]    & $2.5^{+0.9}_{-0.8}$   &  $1.0^{+0.3}_{-0.2}$  &  $0.6^{+0.3}_{-0.2}$ &  $0.92^{+0.30}_{-0.28}$   \\

    $\xi$     & [$\rm{erg~cm~s^{-1}}$]  & $<2.1$   &  $<0.4$  &  $<0.2$ &  $<1.9$  \\

    $\chi^2$/d.o.f.   &   & 128.3/116 &  385.1/307  & 370.7/296 &  389.2/315 \\


    
       \hline
       
    $F_{\rm{AGN,0.5-10}}$   &  [$10^{-11}~\rm{erg~s^{-1}~cm^{-2}}$]
    &$1.6\pm0.7$     &$0.91\pm0.06$ & $1.14\pm0.08$& $1.26\pm0.08$  \\ 
    
    $L_{\rm{AGN,0.5-10}}$
    
    & [$ 10^{41}~\rm{erg~s^{-1}}$]
    & $7\pm3$     &$3.8\pm0.2$     &$4.6\pm0.3$     &$5.2\pm 0.3$  \\


    $F_{\rm{jgal,0.5-10}}$   &
    [$10^{-13}~\rm{erg~s^{-1}~cm^{-2}}$]
    & $3.30\pm0.10$    &  $3.40\pm0.08$     &  $3.50\pm0.16$ &  $3.40\pm0.23$  \\
    
    $L_{\rm{jgal,0.5-10}}$  & [$ 10^{40}~\rm{erg~s^{-1}}$]
    & $1.36\pm0.04$& $1.40\pm0.04$ &$1.44\pm0.06$ &$1.40\pm0.09$  \\

  $Refl$   &   &$ 0.03\pm0.02  $     &$ 0.06\pm0.01 $               & $ 0.05 \pm 0.01 $             & $   0.050\pm0.007$  \\

    

     

    $F_{\rm{AGN,2-10}}$ & [$10^{-11}~\rm{erg~s^{-1}~cm^{-2}}$]
    &$1.03\pm0.20$     &$0.70\pm0.03$      & $0.84\pm0.04$     & $0.90\pm0.02$  \\ 

    $L_{\rm{AGN,2-10}}$  & [$ 10^{41}~\rm{erg~s^{-1}}$]
    & $4.2\pm0.6$        &   $2.9\pm0.1$      &$3.4\pm0.2$        &$3.74\pm0.08$  \\


    $F_{\rm{jgal,2-10}}$   &
    [$10^{-13}~\rm{erg~s^{-1}~cm^{-2}}$]
    & $1.1\pm0.05$         &$1.10\pm0.02$         &$1.1\pm0.05$            &$1.10\pm0.08$  \\
        
    $L_{\rm{jgal,2-10}}$  & [$ 10^{39}~\rm{erg~s^{-1}}$]
    & $4.5\pm0.2$   &  $4.5\pm0.1$ &   $4.5\pm0.2$ &   $4.5\pm0.3$  \\





$F_{\rm{AGN,0.5-2}}$   &  [$10^{-12}~\rm{erg~s^{-1}~cm^{-2}}$]
&   $6.4\pm4.3$    &$2.1\pm0.4$           &$3.0\pm0.5$               &$3.6\pm0.5$   \\

$L_{\rm{AGN,0.5-2}}$  & [$ 10^{41}~\rm{erg~s^{-1}}$]
& $2.6\pm1.8$              &$0.9\pm0.2$ &   $1.2\pm0.2$ &   $1.5\pm0.2$  \\


$F_{\rm{jgal,0.5-2}}$   &  [$10^{-13}~\rm{erg~s^{-1}~cm^{-2}}$]
& $2.2\pm0.1$   &$2.3\pm0.1$ &$2.4\pm0.1$ &$2.3\pm0.2$   \\

$L_{\rm{jgal,0.5-2}}$  & [$ 10^{39}~\rm{erg~s^{-1}}$]
& $9.1\pm0.4$  &$9.5\pm0.4$ &$9.9\pm0.4$ &$9.5\pm0.4$  \\

    \hline

\end{tabular} 
          
       \end{table*}

\begin{table*}
  \centering
  \caption{Results from fits to the {\it Chandra} spectrum of the nucleus between 0.5 and 10~keV. The models are:   PexTH: \texttt{tbabs(zpcfabs*pexmon + mekal)};  PexiTH: \texttt{tbabs(zpcfabs*absori*pexmon + mekal)}. The photon index of the power law has not been allowed to vary below 1.1. All fluxes and luminosities are unabsorbed. The flux of the AGN component has been corrected by a factor of 5, considering that the central (excluded) region contains 80  per cent of the PSF.
The errors on AGN fluxes are the average errors of the reflected and direct component.  
  }
 \label{tab:chandra_results}
\begin{tabular}{llll} 
	  \hline
 Parameter & Unit  &  PexTH  &     PexiTH   \\
          \hline
          
      $N_{\rm{H,high}}$ &  [$10^{22}~\rm{cm^{-2}}$]  &  $12^{+4}_{-3}$  &    $   24_{-11}^{+30}  $     \\    
    
  Cov-frac  &   &   $0.79  \pm 0.06 $    &      $0.6  \pm 0.1 $     \\ 

  $\Gamma$ &    &   $  <1.5 $   &     $ 1.13 ^{+0.24}_{-0.03} $       \\

     $N_{\rm{H,low}} $ & [$10^{22}~\rm{cm^{-2}}$] &   - &    $ 7_{-4}^{+5}  $     \\

  $\xi $    &[$\rm{erg~cm~s^{-1}}$]  &   -  &     $ 746 _{-470}^{+4220}  $    \\

          $T $ & [keV] &  $ 0.62_{-0.05}^{+0.05}  $   &      $ 0.71 _{-0.06}^{+0.07}  $     \\

  $\chi^2/$d.o.f. &  &  90.2/69  &     74.8/67   \\   

\hline





$F_{\rm{AGN,0.5-10}}$   & [$10^{-12}\rm{erg~cm^2~s^{-1}}$] & - & $10 \pm 6$ \\

$L_{\rm{AGN,0.5-10}}$      & [$10^{41}\rm{erg~s^{-1}}$]  &- & $4\pm 2$ \\


$F_{\rm{th,0.5-10}}$   & [$10^{-14} \rm{erg~cm^2~s^{-1}}$] & - &  $ 3.8 \pm 0.4 $   \\

$L_{\rm{th,0.5-10}}$      & [$10^{39} \rm{erg~s^{-1}}$]  &- & $ 1.5\pm 0.2 $ \\



\hline

$Refl$  & & - & $0.05\pm0.05$  \\



$F_{\rm{AGN,2-10}}$   & [$10^{-12} \rm{erg~cm^2~s^{-1}}$] & - &  $ 8 \pm  5 $   \\

$L_{\rm{AGN,2-10}}$   & [$10^{41} \rm{erg~s^{-1}}$]&- & $ 3.2 \pm 2.1 $  \\


$F_{\rm{th,2-10}}$   & [$10^{-15} \rm{erg~cm^2~s^{-1}}$] &- & $ 1.4 \pm 0.3 $     \\

$L_{\rm{th,2-10}}$ & [$10^{37} \rm{erg~s^{-1}}$]  - & & $5.8\pm 1.2$\\




\hline 
$F_{\rm{AGN,0.5-2}}$   &[$10^{-12} \rm{erg~cm^2~s^{-1}}$]  & - &  $1.7  \pm   0.6 $   \\

$L_{\rm{AGN,0.5-2}}$   & [$10^{40} \rm{erg~s^{-1}}$]  & - &  $7 \pm2 $ \\

$F_{\rm{th,0.5-2}}$   & [$10^{-14} \rm{erg~cm^2~s^{-1}}$]  & - &  $ 3.7 \pm0.5  $   \\

$L_{\rm{th,0.5-2}}$   & [$10^{39} \rm{erg~s^{-1}}$]  &   - & $1.5\pm0.2$ \\

\hline

\end{tabular} 
          
       \end{table*}

\subsection{Emission from the extended jet and the galaxy}
\label{sec:disc-jet}

We have used the {\it Chandra} observation to explore the X-ray morphology and spectra of the extended emission in NGC~1052, including a comparison with the radio jets. VLA  1.4 GHz (Fig. \ref{fig:chandra_soft_radio} of this paper) and MERLIN images (Fig. 3 of \citealt{kadler2004}) show
 a pair of  hotspots for both the east and west lobes, which may be due to multiple jet ejection activity as in episodic radio galaxies. While this is more common in luminous sources \citep{nandi2019}, extended radio lobes on kpc scales are not common in LINER, with only a few examples in the literature \citep{wierzbowska2012}. The double structure seen at 1.4~GHz in NGC~1052 needs higher resolution data to be confirmed, but the current images make this source stand out as one of few candidates for episodic jet activity in LINER.  We also note that the radio morphology of the lobes is asymmetric,  with the east lobe showing a bent structure towards the south.  

From the X-ray images, we find that there is soft, extended emission along the same direction as the radio structure. An X-ray hotspot located at the outer edge of the eastern radio lobe is clearly seen below 2~keV. Its position indicates emission at the impact point between the jet and the environment. Similarly, there is some X-ray emission along the south-west edge of the western radio lobe. Some X-ray radiation is also seen below 1~keV at a position coincident with the inner radio hotspot on the eastern side, but a deeper X-ray observation would be needed to confirm a possible association with the radio hotspot. 

Fig.~\ref{fig:chandra_soft_radio} shows that there is also extended emission from the galaxy that does not exactly overlap with the radio jets. Diffuse emission distributed over large scales is seen between 0.5 and 1~keV, while it is more compact between 1 and 2~keV. Moreover, we note that only the nucleus is detected above 2~keV. The different spatial scales seen  between $0.5-1$~keV and $1-2$~keV suggest two ionising sources of different nature. A compact high-energy source ionises the gas emitting  between 1 and 2~keV, while an extended and lower energy source ionises the gas seen between  0.5 and 1~keV.  The first ionising source can be identified as the central nucleus, while the second one is instead distributed over larger scales consistent with the scales of the ISM. 
This ionising radiation may come from discrete sources unresolved in the X-rays. 
  Evidence for discrete sources in the central 1 kpc region has been found in high-resolution NIR images \citep{fernandez-ontiveros2010}. These sources, identified as star forming clusters, could be the reminiscence of a minor merger event reported in
  \cite{vanGorkom1986},  connected with recent ($<1$Gyr) star formation activity. However, it is unlikely that this limited star-formation activity is sufficient to ionise the extended X-ray emission.

We fitted {\it Chandra} spectra of the regions containing emission from the extended  jet and the galaxy using a phenomenological model composed of a thermal component and a power law. The thermal model represents a hot ($T\sim0.6$~keV) optically thin plasma. The temperature is consistent with shocks between the jet and the ISM, as well as emission from highly star forming regions. A similar temperature was reported in \cite{kadler2003} and \cite{kadler2004}, who pointed out its consistency with emission from jet-triggered shocks.  The spatial distribution of emission in the {\it Chandra} images supports this scenario, as discussed above. Jet shocks convert a few percent of the kinetic energy of the jet into X-ray emission. From \cite{kadler2003}, the kinetic power of the jet is $\sim 5 \times 10^{41}~ \rm erg~s^{-1}$.
The luminosity of the thermal component in the fits to the spectra of the extended emission is $\sim8.6 \times 10^{38}~\rm{ erg~s^{-1}}$ (Table~\ref{tab:chandra_jgalfits}), which corresponds to  $< 0.2$~per~cent of the jet kinetic energy, lower than the upper limits obtained in the previous {\it Chandra} observation ($< 2.5$ per cent,  \citealt{kadler2003}). Our estimate is also an upper limit since there is most likely some contribution to the thermal emission from gas not associated with the jet.

The power-law component in our spectral fit has a less clear interpretation.
Emission from the extended jet is a likely candidate given that  X-ray emission is observed along the direction of the radio lobes. However, emission from diffuse gas and weak point sources, such as X-ray binaries, may also contribute to this spectral component. 
If the jet contributes significantly to the power law, the steep photon index in the fit indicates that the origin may be synchrotron emission. The limited count statistic does not allow us to disentangle the possible jet emission from the environment in the spectral fitting.

Finally, we note that the {\it XMM-Newton} spectra may show evidence for emission from the galaxy in addition to the two components discussed above. Our analysis of the last {\it XMM-Newton} spectrum from 2009 (which has the best statistics) indicates a small contribution from an additional thermal component with a temperature of $1.3\pm0.2$~keV, which might be associated with a diffuse background or discrete background sources. Such a contribution is plausible since the X-ray image between 1 and 2 keV in Fig. \ref{fig:chandra_soft} show two background sources located south of the nucleus outside the jet plus galaxy regions, as well as some extended emission.

\subsection{The core}
\label{sec:disc-core} 

We have analysed five spectra of the core of NGC~1052 (four from {\it XMM-Newton} and one from {\it Chandra}), which probe the properties of the source on time-scales from months to years. 
The best fitting model includes an absorbed primary power-law, contributions from the extended jets and the galaxy in the soft X-ray band, as well as a narrow iron line at 6.4 keV and associated Compton reflection from distant material.

The results on each of these components are discussed below.
The spectral components describing the emission associated with the extended jets and galaxy in the {\it XMM-Newton} spectra were determined from the fits to the {\it Chandra} spectra of the extended emission discussed above. We allowed only the normalisation of this extended emission to vary in the fits to the  {\it XMM-Newton} spectra, finding that it accounts for about half of the observed flux in the $0.5-2$~keV band.

\subsubsection{Primary continuum}

We find that the fluxes of the power-law components in the five observations are mostly consistent within the uncertainties, with the exception of the second {\it XMM-Newton} observation, which shows evidence for a significantly lower flux ($\sim30$\% lower than in the fourth observation).
Previous studies of the source have also found significant flux variability \citep{hernandez-garcia2013,connolly2016,osorio2019}. Compared to these works, we have somewhat larger uncertainties on the fluxes, which is due to the more complex model adopted. 
This will be discussed in more detail in Sect. 4.4.

We found that the primary power law makes up $\sim95$ per cent of the unabsorbed flux above 2~keV (Table~\ref{tab:xmm_fits}), showing that accretion onto the SMBH dominates the emission at these energies. \cite{terashima2002} discussed the possibility that galactic emission components may contribute to the spectra of LINER above 2~keV, producing a power law and/or thermal radiation at high energies. These contributions, if relevant, would complicate the spectral analysis. However, the {\it Chandra} images of NGC~1052 show extended emission primarily in the soft-X-ray band. Both the spectral analysis and the images thus show that the bulk of the radiation above 2~keV originates from the nucleus itself.

The photon indices from the {\it XMM-Newton} fits are in the range 1.34$\pm0.09$ to 1.8$\pm0.3$ (Table~\ref{tab:xmm_fits}). These values are lower than in standard AGN, which have photon indices typically in the range between 1.8 and 2.2 \citep{yang2015}.
There are several possible explanations for the low photon index. One possibility is that emission from jets contribute to the spectrum. This may be due to a residual contribution from the extended jets or, more plausibly, a compact jet in the innermost regions.
Indeed, there is evidence for an unresolved jet  located in the nucleus of NGC1052 \citep{baek2019}.
This is in line with the fact that compact radio cores, which are common in LINER \citep{ho2008,baldi2019}, may be indicative of unresolved jets \citep{markoff2005,markoff2008}.
The jet scenario is also supported by the harder-when-brighter spectral variability reported in {\it Swift}/XRT observations of this source  \citep{connolly2016}. This kind of variability is commonly associated with synchroton self-Compton emission in blazars, with the increase in luminosity being connected to a hardening of the electron population (e.g., \citealt{ghisellini2009}). This scenario has also been discussed as an explanation for  the harder-when-brighter behaviour in  the LLAGN NGC~7213 \citep{emmanoulopoulos2012}. 
By analogy with the jet scenario, the same kind of spectral variability would be expected due to Comptonisation of synchrotron seed photons in a corona, as discussed for LLAGN in  \cite{connolly2016}. In this case, the synchrotron emission may originate from the corona itself or the ADAF.  
Finally, an alternative possibility is that the flat X-ray continuum  is due to thermal radiation from the ADAF, as argued in \cite{guainazzi2000}. These authors analysed  previous observations of NGC~1052 from {\it BebboSAX}, finding that a flat power law fits the spectrum equally well as a thermal Bremsstrahlung model with a temperature of $\sim150$~keV. 

\subsubsection{Absorption and reflection}

We have investigated deviations from the power-law radiation considering material around the central engine. This material can have the effect to both absorb soft X-ray photons and reprocess the higher energy radiation to produce fluorescent lines and Compton reflection.
Our best-fit model comprises absorption from both a partially covering cold medium and a warm absorber. While the first one is most likely associated with cold material in the torus, the latter may be explained as material along the line of sight in the galaxy or as outflows. Recent optical spectroscopy of NGC~1052 show evidence of possible outflows, preferentially seen in [O I], in agreement with the second scenario \citep{cazzoli2018}. However,  our X-ray analysis shows an ionisation degree consistent with the neutral state.

  
The column densities of the clumpy and homogeneous absorbers  are well determined from the spectral fitting (i.e. their uncertainties do not overlap with each other) and they are not degenerate with the other main continuum parameters like the photon index.  
We can conclude that the primary radiation is convolved by cold absorption in non-homogeneous material of high density ($\sim9 \times 10^{22}\rm{ cm^{-2}}$)  and covering fraction of $ \sim$80-90\% (all parameter constraints rule out 100\% coverage) and in a low density ($\sim1 \times 10^{22} \rm{cm^{-2}}$) homogeneous gas. The column densities of the two absorbers are roughly constant across the observations with good spectral quality, with significantly different values found only in the two spectra with the lowest count statistics.
We note that some previous studies have reported variations in the absorbing column (e.g., \citealt{hernandez-garcia2013} and \citealt{osorio2019}). As discussed in section 4.4., these discrepancies are due to differences in the models used in the different studies. 

The iron $K_{\alpha}$ line at 6.4~keV is clearly detected in all spectra.
The line is well modelled by a narrow gaussian profile, which
means that reflection occurs in material distant from the central engine, most likely the torus. The equivalent width measurements  of 130-180~keV (see Table \ref{tab:xmm_fits_norefl}), together with the column density estimates for the absorbers, suggest that the source is not heavily absorbed.
Analysing the spectra considering iron line and Compton reflection in a self consistent way, we found that the reflection comprises 3-6~\% of the unabsorbed AGN flux. The photon index of the primary radiation becomes slightly steeper when adding Compton reflection because the fit takes into account separately the (harder) reflection contribution.

We tested for the presence of relativistic reflection from the accretion disk using both the \texttt{laor} model to represent the iron $K_{\alpha}$ line distorted by relativistic effects, and \texttt{relxill} to represent the full reflection spectrum. 
The relativistic contribution does not introduce any fit improvement with respect to reflection in material distant from the SMBH.
 In our analysis, the constraint on the inner radius on the first {\it XMM-Newton} spectrum is $r_{\rm{in}}<370$~$r_{\rm{g}}$ which does not support relativistic effects. 
 This is expected in the ADAF scenario, where the standard disk is truncated far from the SMBH.
The result is also consistent with previous studies of NGC~1052 and LINER surveys \citep{osorio2019, hernandez-garcia2014, gonzalez-martin2009b}, where non-relativistic reflection was used to fit the data. 
Our constraint on the inner radius is also larger  than the $<$45 $r_{\rm{g}}$ obtained in the analysis of {\it Suzaku} data by \cite{brenneman2009},
who reported a broad line with an
EW of 185 eV.
We consider several scenarios to explain this in Section \ref{differences}.

\subsubsection{The ADAF nucleus}

To understand the accretion mode of the nucleus in NGC~1052 we computed the Eddington ratio as $\log{ ( L_{\rm{bol}}/L_{\rm{edd}}}$), where $L_{\rm{bol}}$ is the bolometric luminosity and $L_{\rm{edd}}$ is the Eddington luminosity. Given the BH mass of $1.5\times10^{8}~{\rm M}_{\odot}$ \citep{woo2002}, the Eddigton luminosity is $1.8\times10^{46}~\rm{erg~s^{-1}}$.
  The bolometric luminosity estimated from the integration of the Spectral Energy Distribution (SED) is $\sim7\times10^{42}\rm~erg~s^{-1}$ \citep{fernandez-ontiveros2019}, which gives an Eddington ratio of -3.4,  consistent with accretion in a radiatively inefficient mode. 
  In this case, the optically thick, geometrically thin accretion disk is expected to be truncated and replaced by an ADAF at the centre \citep{quataert1999}. 
  
Similar targets classified as LLAGN (and with Eddington ratios similar to NGC~1052) have a transition radius between the standard disk and the ADAF around 100 Schwarzschild radii\footnote{The Schwarzschild radius is $2 \times r_g$}, as discussed in \cite{yuan2004}. However, modelling of the SED of NGC~1052 with an ADAF-scenario has yielded mixed results. While \cite{yuan2009} finds that the source is well described by an ADAF-dominated model, the ADAF+jet model of \cite{yu2011} significantly over-predicts the UV flux, which is instead better described by a standard disk extending to the innermost stable circular orbit. More recently, \cite{reb2018} analysed the sub-arcsec  SED and found that the standard disk is truncated at a radius $> 26\ {r_{\rm g}}$. The scenario of a truncated disk is also supported by our analysis, which shows that the X-ray spectra lack signs of relativistic reflection.

\subsection{Comparison with previous works}
\label{differences}

Here we discuss our findings regarding the nucleus and surrounding regions in NGC~1052 in relation to previous studies of the X-ray properties of the source. Our analysis of the extended emission includes the first comparison between images from the 2005 {\it Chandra} observation and VLA observations.
The X-ray images show diffuse emission extended along the radio lobes, as well as hotspots associated with the edges of the lobes. This confirms previous results obtained by \cite{kadler2004} with a much shorter {\it Chandra} observation. While the diffuse X-ray emission has a similar extent at the radio lobes, the detailed structures do not correlate well, which indicates that the X-ray and radio emission trace different parts of the particle population in the extended jets. Indeed, differences between X-rays and radio are commonly seen in jets, including the nearby well-observed cases of Cen~A and M87 \citep{kraft2002,perlman2005}. In addition to offsets between X-ray and radio hotspots in the jets of both these sources, the Cen~A jet is also narrower in X-rays than in radio \citep{kraft2002}.

 We characterised the spectrum of the circum-nuclear emission,  disentangling it from the nucleus. The area selected for this study includes the X-ray emission from the radio lobes as well as part of the galaxy, but excludes the central core.
An estimate of the circum-nuclear spectrum from the same data was recently reported in  \cite{osorio2019},
based on an annular extraction region
beween 3$\arcsec$ and 25$\arcsec$.
They find a thermal component consistent with our results, but a significantly flatter power law ($0.7\pm 0.2$ compared to our value of $2.3\pm 0.2$). This discrepancy is likely due to differences in the extraction region. Since ~10\% of the emission from a point source is outside 3~arcsec at 8.5 keV\footnote{http://cxc.harvard.edu/proposer/POG/html/chap4.html}), it is plausible that the harder photon index reported by \cite{osorio2019} is affected by a small contribution from the bright and hard nucleus.
Our extraction region excludes the central 6$\arcsec$ radius, which ensures that there is no emission from the nucleus contributing to the spectrum. 
The different circum-nuclear spectrum likely also explains the slightly lower AGN fluxes in \cite{osorio2019}, considering that part of the nucleus may be included in the circum-nuclear model.
From our analysis, the AGN power law of the spectra with good statistics are also somewhat harder (all spectra except for the first, lowest quality spectrum from \textsl {XMM-Newton}), again probably due to the new  characterisation of the circum-nuclear spectrum.

The overall model for the nucleus adopted in this work contains an absorber composed by partially covering material and a warm absorber, with physical parameters broadly consistent with previous works, considering the uncertainties.
We thus confirm previous findings from \cite{osorio2019} and \cite{brenneman2009}, which show that non-homogeneous cold absorption in combination with a lower density absorbing material fully covering the source fit the data better than a simple cold uniform absorber.
The model used to fit the {\it XMM-Newton} spectra also requires  a small contribution from reflection in material distant from the central SMBH.
Although the overall model  provides an acceptable fit to the broad-band spectra, we found indications for an additional thermal component that improves the fit in the soft X-rays. 
We argue that this might be the result of a small contribution from background radiation outside the two jet+galaxy regions.

Our analysis reveals weak evidence for variability in the best-fit spectral parameters (see Table 3). The first {\it XMM-Newton} observation and the {\it Chandra} observation show higher values of the column densities of the absorbers, but these two observations
also have the lowest spectral quality. There is also marginal evidence for variability in $\Gamma$. Previous studies of NGC~1052 have revealed significant variations in hardness ratios using observations from both {\it XMM-Newton} \citep{hernandez-garcia2013} and {\it Swift} \citep{connolly2016}. The analysis of \cite{hernandez-garcia2013} mainly attributed this to changes in the absorbing column, while\cite{connolly2016} favoured a varying photon index. In comparison to these works, we have considered a more comprehensive physical scenario with a larger number of free parameters. From this analysis we cannot attribute the spectral variability to a specific parameter or component of the model.

Using a model that is more similar to ours, \cite{osorio2019}
 find small variability in $\Gamma$  and the column density of the clumpy absorber between the three best {\it XMM-Newton} observations (the ones from 2006-Jan, 2009-Jan, and 2009-Aug). The tighter constraints on these parameters from \cite{osorio2019} arise from the fact that some of the other model parameters were kept constant or tied across the observations, whereas we have allowed all the parameters in Table \ref{tab:xmm_fits} to vary independently. We thus conclude that the spectral variability between these three observations is small enough that it does not result in significant differences between the best-fit parameters when allowing all of them to vary.

Our analysis does not reveal a significant relativistic iron line as was reported for the previous {\it Suzaku} data \citep{brenneman2009}.
 The  relativistic line in the {\it Suzaku} spectrum  had an EW of 185 eV and gave a constraint on the inner disk radius of $<$45 $r_{\rm{g}}$.  We consider different scenarios to understand this difference with respect to our results.
 We start by noting that it cannot be explained by differences in the spectral quality because the best {\it XMM-Newton} spectra have 35000 counts in the X-ray band, while the {\it Suzaku} data have 39000 total counts from 0.5 to 9~keV. However, we note that the first {\it XMM-Newton} spectrum has a hint of a red wing of the line, although it does not clearly stand out from the background noise (see Fig. \ref{fig:xmm_allplots}, panel a).  The possibility of a relativistic contribution to the line in the same spectrum was noted in \cite{ros2007}, who also discussed a possible connection with a jet ejection that would have occurred in 2001, as found in radio data, right before the {\it XMM-Newton} observation.
In the case of the {\it Suzaku} data, which does not show any Compton reflection associated with the relativistic line, it was also proposed that the line emission may originate in the base of the jet \citep{brenneman2009}. 
 Line emission from the base of the jet is not expected to arise together with Compton reflection and could in principle be variable on  time scales of years, as discussed in \cite{ros2007}.
 However, we cannot verify this because of the limited spectral quality of the {\it XMM-Newton} spectrum of 2001 (it has only 4000 counts).
 
The detection of relativistic line emission in the {\it Suzaku} spectrum of 2007 could alternatively be explained, at least in part, as due to a different way to model the continuum in \cite{brenneman2009}. The continuum model for the {\it Suzaku} spectrum was determined from fits that ignore the spectral band beween 3~keV and 7~keV to exclude the iron line region. The 3 to 7~keV band was then reintroduced when fitting the relativistic line, which may lead to an underestimate of the continuum.
Indeed, it is extremely rare to find relativistic lines extended down to 3~keV; even in luminous AGN the most extreme cases extend down to 4~keV (see for instance \cite{nandra2007} or \cite{risaliti2013} for a more recent analysis).
Our continuum model for the {\it XMM-Newton} spectra differs from the one used for the {\it Suzaku} spectra in that it includes the emission components inferred from the {\it Chandra} spectrum of the jet and the galaxy.  The inclusion of these soft X-ray components affects the other parameters of the fit (including the absorbers and the power-law), thus modifying the continuum model also at high energies. 
Consequently, the more detailed continuum characterisation may affect the conclusions regarding the presence of a relativistic Fe line.

Taken at face value, the relativistic iron line found in {\it Suzaku} but not detected in the {\it XMM-Newton} observations, can finally be interpreted as the genuine result of some variability in the innermost regions of the accretion disc.
We checked if different pictures besides the jet ejection event can explain variability in the relativistic line, considering in particular instabilities in the accretion disk (e.g. \citealt{McHardy2006} and \citealt{merloni2003}).
We thus compare the fluxes measured in the {\it Suzaku} observation of July 2007 with the closest {\it XMM-Newton} observations, obtained in January 2006 and January 2009. The latter have fluxes between 2 and 10 keV of $\sim7.0\pm0.3$ and $\sim8.4\pm0.4$ $10^{-12}$ erg$\rm{~cm^{-2}~s^{-1}}$, respectively, marginally lower than the measurement from {\it Suzaku} of 9$\pm1\times10^{-12}$ erg $\rm{cm^{-2}~s^{-1}}$ reported in \cite{brenneman2009}. We conclude that there is no significant continuum flux variability in support of accretion disk transitions.

\section{Conclusions}
We have studied one {\it Chandra} observation and four {\it XMM-Newton} observations of NGC~1052 in order to explore its innermost structure along the scales of the radio jets and  to determine the nature of the emission. Our main conclusions are the following:

\begin{itemize}
\item  The {\it Chandra} images reveal diffuse emission below 1~keV extended along the radio lobes. 
However, the diffuse X-ray emission does not correlate well with the radio contours, suggesting that the X-ray and radio emission trace different parts of the particle distribution of the extended jets. In addition,  

 there is a faint hotspot at the border of the eastern radio lobe and also some emission along the southern edge of the western lobe. 
 This can be explained as emission from shocks between the extended jets and the environment. 

\item The {\it Chandra} spectrum extracted from the regions containing emission from the extended jets and galaxy can be described by a thermal  component and a power law. While the main contribution to the thermal component are the X-ray hotspots coincident with the radio lobes, the power law likely represents the emission from the jets, with a possible contribution from other weak sources in the galaxy.  A comparison with the FIR luminosity shows that star formation is expected to make a very small contribution to the X-ray emission.

\item 
  
From the images, we see that the main emission above 2 keV originates from the nucleus. The spectral analysis shows that the continuum from the nucleus is well modelled by a hard power law ($\Gamma \sim 1.4$). This can be interpreted as the emission from a hot ADAF, a compact jet, or Comptonisation in a corona, where the seed photons in the latter case may be synchrotron photons from the ADAF or the corona itself.

\item The continuum spectrum is absorbed by a complex system comprising a high density ($N_{\rm H,high} \gtrsim$ $9\times10^{22} \rm{cm^{-2}}$) patchy torus (covering fraction $\sim90$ per cent) and a diffuse absorbing structure with a lower density ($N_{\rm H,low} \lesssim$ $2\times10^{22}\rm{cm^{-2}}$), likely associated with diffuse gas along the line of sight.
  
\item A narrow iron line at 6.4~keV is significantly detected in all spectra of the nucleus, providing clear evidence of accretion onto the central SMBH. The associated Compton reflection from distant material accounts for 3-6~\% of the total AGN flux between 2 and 10~keV. 

\item The spectra lack a relativistic line at 6.4~keV. This means that the inner edge of the accretion disk is too far away from the central SMBH for relativistic effects to be detected. The spectral analysis constrains the inner radius to $< 370~r_{\rm g}$, considering a continuum composed of a power law and Compton reflection. 
This is in agreement with radiatively inefficient accretion in the form of an ADAF hosted in a truncated accretion disk.

 \end{itemize}

\begin{acknowledgements}

  We thank the referee for the constructive comments, which have greatly improved the paper.
 JL and SF thank the Knut \& Alice Wallenberg Foundation and the  Swedish National Space
Board.
SN acknowledges support by the Science \& Engineering Research Board, a statutory body of Department of Science and Technology (DST), Government of India  (FILE NO. PDF/2018/002833).
This research has made use of data obtained from the {\it Chandra} Data Archive and software provided by the {\it Chandra} X-ray Center (CXC) in the application packages \texttt{CIAO} and \texttt{CHIPS}.
This research has made use of the NASA/IPAC Extragalactic Database (NED),
which is operated by the Jet Propulsion Laboratory, California Institute of Technology,
under contract with the National Aeronautics and Space Administration.

\end{acknowledgements}




\bibliographystyle{aa}
\bibliography{bibliography} 



\begin{appendix}
\section{Fits with simple continuum models}
  
This section reports fits parameters, fluxes and luminosities estimated from the {\it XMM-Newton} nuclear spectra fitting the spectra with simple models which ignore Compton reflection.  
The fit results are presented in Table A.1. The first two entires in the table are the P-JGAL and PG-JGAL models introduced in section 3. The third model was obtained by modifying PG-JGAL model introduced in Sect. 3, to take into account extra absorption as we did for the fits reported in Sect. 3. For this, we used a warm absorber (\texttt{absori}), giving the total model \texttt{tbabs(zpcfabs*absori*pow + zgauss + mekal + pow)}.  The model provides an acceptable fit to all four spectra and the main spectral parameters are broadly consistent with the ones reported in Sect. 3.


   



\renewcommand{\arraystretch}{1.5}
\begin{table*}
  \centering
  \caption{Results of fits to the {\it XMM-Newton} spectra between 0.5 and 10~keV. The models are P-JGAL: \texttt{tbabs*(zpcfabs*pow+mekal+pow)}.  PG-JGAL:  \texttt{tbabs*(zpcfabs*pow+GAUSS+mekal+pow)}. PGI-JGAL:  \texttt{tbabs*(zpcfabs*absori*pow+GAUSS+mekal+pow)}. The last two \texttt{mekal+POW} components in all three models represent the extended emission from the jet and galaxy (see Table~\ref{tab:chandra_jgalfits} and text for details). All fluxes and luminosities are unabsorbed.  N$_{\rm H,high}$ and  N$_{\rm H,low}$  are the column densities of the \texttt{zpcfabs} and \texttt{absori} models, respectively,  and $\xi$ is ionisation parameter of \texttt{absori}. EW is the equivalent width of the narrow iron line. }
	\label{tab:xmm_fits_norefl}
    \begin{tabular}{l|l|r|r|r|r|r}\\
    	  \hline
	  & & & \multicolumn4c{OBSID}  \\	  
  Model &    Parameter    &   Unit &     0093630101  &  0306230101   &   0553300301 & 0553300401   \\
\hline

 P-JGAL  &      &           &     &    &    \\
 &  $\chi^2$/d.o.f.   &  &231.0/119  & 571.1/310  & 499.0/299 &   563.0/318  \\

\hline
          PG-JGAL    &   &               &     &    &  &  \\

 &      $N_{\rm{H,high}}$ &  [$10^{22}~\rm{cm^{-2}}$] & $16^{+2}_{-2} $     & $7.4^{+0.5}_{-0.5} $ & $7.7^{+0.5}_{-0.5} $ & $8.2^{+0.4}_{-0.4} $ \\
   
          &  Cov-frac  &   & $0.90^{+0.02}_{-0.02} $        & $0.88^{+0.01}_{-0.01} $ & $0.920^{+0.008}_{-0.008} $ & $0.930^{+0.005}_{-0.005} $ \\
   
 &  $\Gamma$ &   & $1.51^{+0.09}_{-0.09} $            & $1.05^{+0.07}_{-0.07} $ & $1.24^{+0.06}_{-0.06} $ & $1.32^{+0.06}_{-0.06} $ \\

  &   EW & [eV]  & $180^{+65}_{-65} $                     & $183^{+25}_{-25} $ & $150^{+26}_{-26} $ & $152^{+22}_{-22} $ \\

  & $\chi^2$/d.o.f.     &        & 211.8/118  & 430.2/309  &  406.2/298 & 430.7/317  \\

 \hline  

 PGI-JGAL   &     &    &   &   &  & \\

   &    $N_{\rm{H,high}}$    & [$10^{22}~\rm{cm^{-2}}$]  & $25^{+6}_{-5} $   
    &  $ 10^{+1}_{-1}$            &  $8.6^{+0.7}_{-0.6}$  &  $9.0^{+0.6}_{-0.5}$   \\   
    
 &   Cov-frac     &   & $0.86^{+0.04}_{-0.06} $   
    &  $ 0.84^{+0.02}_{-0.02}$            &  $0.91^{+0.01}_{-0.01}$  &  $0.92^{+0.01}_{-0.01}$   \\  

 &   $\Gamma$ &   & $1.72^{+0.33}_{-0.31} $   
    &  $ 1.27^{+0.02}_{-0.02}$            &  $1.38^{+0.09}_{-0.08}$  &  $1.46^{+0.07}_{-0.07}$   \\

  &    $N_{\rm{H,low}}$ & [$10^{22}~\rm{cm^{-2}}$]  & $2.34^{+0.87}_{-0.79} $   
    &  $ 0.99^{+0.27}_{-0.21}$            &  $0.59^{+0.35}_{-0.22}$  &  $0.90^{+0.27}_{-0.29}$   \\

 &   $\xi$     & [$\rm{erg~cm~s^{-1}}$] & $<2.5 $   
    &  $ <0.6$            &  $<0.2$  &  $<2.5$   \\

 &   EW    & [eV]  & $136^{+61}_{-61} $   
    &  $ 169^{+24}_{-24}$            &  $146^{+23}_{-23}$  &  $148^{+22}_{-22}$   \\

 &   $\chi^2$/d.o.f.   &   & 129.7/116   & 385.7/307    & 370.4/296 & 394.5/315  \\ 

 
        
        
        
     \hline   

\end{tabular} 
          
       \end{table*}

\end{appendix}


\end{document}